\documentclass[aps,prx,twocolumn,amsmath,amssymb,floatfix]{revtex4-1}

\usepackage{dcolumn}
\usepackage{bm}
\usepackage{epsfig}
\usepackage{epstopdf}
\usepackage{graphicx}
\usepackage{grffile}
\usepackage[english]{babel}
\usepackage{setspace}

\usepackage{bm}
\usepackage{sidecap}
\usepackage{mathptmx}
\usepackage{amsmath}
\usepackage{txfonts}
\usepackage[colorlinks,citecolor=blue,linkcolor=purple]{hyperref} 
\usepackage[usenames,dvipsnames,svgnames]{xcolor}
\usepackage{array}
\usepackage{float}

\usepackage{lipsum}
\usepackage{natbib}
\usepackage{lineno}

\usepackage{soul}
\usepackage{ulem}

\begin{document}
\renewcommand{\thefigure}{\arabic{figure}}
\def\be{\begin{equation}}
\def\ee{\end{equation}}
\def\ber{\begin{eqnarray}}
\def\eer{\end{eqnarray}}

\def\kv{{\bf k}}
\def\bfr{{\bf r}}
\def\qv{{\bf q}}
\def\pv{{\bf p}}
\def\sigmav{\boldsymbol{\sigma}}
\def\tauv{\boldsymbol{\tau}}
\newcommand{\h}[1]{{\hat {#1}}}
\newcommand{\hdg}[1]{{\hat {#1}^\dagger}}
\newcommand{\bra}[1]{\left\langle{#1}\right|}
\newcommand{\ket}[1]{\left|{#1}\right\rangle}

\title{Large Josephson current in Weyl nodal loop semimetals due to odd-frequency superconductivity}
\date{\today}

\author{Fariborz Parhizgar}\email{fariborz.parhizgar@physics.uu.se}

\author{Annica M. Black-Schaffer}\email{annica.black-schaffer@physics.uu.se}
\affiliation{Department of Physics and Astronomy, Uppsala University, Box 516, SE-751 20, Uppsala, Sweden}
\begin{abstract}
Weyl nodal loop semimetals (WNLs) host a closed nodal line loop Fermi surface in the bulk, protected zero-energy flat band, or drumhead, surface states, and strong spin-polarization. The large density of states of the drumhead states makes WNL semimetals exceedingly prone to electronic ordering. At the same time, the spin-polarization naively prevents conventional superconductivity due to its spin-singlet nature. 
Here we show the complete opposite: WNLs are extremely promising materials for superconducting Josephson junctions, entirely due to odd-frequency superconductivity. By sandwiching a WNL between two conventional superconductors we theoretically demonstrate the presence of very large Josephson currents, even up to orders of magnitude larger than for normal metals. The large currents are generated both by an efficient transformation of spin-singlet pairs into odd-frequency spin-triplet pairing by the Weyl dispersion and the drumhead states ensuring exceptionally proximity effect. As a result, WNL Josephson junctions offer unique possibilities for detecting and exploring odd-frequency superconductivity.
\end{abstract}

\maketitle


\section{Introduction}
In-between a conventional metal with its two-dimensional (2D) Fermi surface and a Weyl semimetal with its 0D Fermi points, we find the Weyl and Dirac nodal loop semimetals (WNLs and DNLs), which have 1D nodal loop Fermi surfaces \cite{Ali-2018, Gao-2018}.
Multiple such materials have recently been both proposed \cite{Chen-2015, Schaffer-2015, Kim-2015, Yu-2015, Bouhon-2017, PRL.107.186806, PRL.121.106402,PRL.121.166802,Fang-2016} and experimentally observed in compounds such as PbTaSe$_2$, ZrSiS, Ca$_3$P$_2$ , CaAgAs \cite{Bian-2016-1, Schoop-2016, Xie-2015, Takane-2018}. 
Away from the nodal loop Fermi surface the dispersion is Weyl-like, completely locking the electron momentum to the electron orbital (for DNLs) or spin (for WNLs) degrees of freedom. As a result, the nodal loop in DNLs possesses a four-fold degeneracy, while WNLs break spin-degeneracy and thus have only a two-fold degenerate loop (crossing non-spin-degenerate valence and conduction bands).
While the explicitly broken spin-degeneracy have so far made the experimental realization of WNLs more demanding, there already exists candidate WNLs. For example, HgCr$_2$Se$_4$ was recently shown using {\it ab-initio} calculations to be a WNL \cite{PRL.107.186806, PRL.121.106402} and experimental probes have also been proposed for how to easily detect the spin-polarization \cite{PRL.121.166802}. Moreover, spin-polarization has experimentally been found in PbTaSe$_2$, making it a likely WNL candidate \cite{Bian-2016-1}. 

The nodal loop Fermi surface results in drumhead surface states at zero energy, whose area is set by the projection of the nodal loop on the surface plane \cite{Bzdusek-2016, Weng-2015, Pezzini-2017}. The origin of this surface state is thus similar to that of the Weyl semimetals, but there only 1D surface arcs are formed due to the lower dimensionality of the bulk Fermi surface. The flat band dispersion of the drumhead states results in a large peak in the surface density of states (DOS). 
As a result, the surfaces become extremely prone to electronic ordering, including superconductivity which has already been discussed as a possibility for systems with surface flat bands \cite{Kopnin-2011, Heikkila-2011, Lothman-2017}. In the bulk of WNLs, a 3D chiral superconducting state has also already been proposed based on both symmetry analysis \cite{Nandkishore-2016} and renormalization group calculations \cite{Sur-2016}. However, the strong spin-polarization of WNLs, and in some cases even a complete spin-polarization of the drumhead surface states, has been assumed to prohibit any spin-singlet superconductivity, and in particular including proximity effect from conventional superconductors (SCs) \cite{Wang-2017}.

The incompatibility of spin-polarization and conventional (spin-singlet, $s$-wave) superconductivity has actually been remedied in a few other cases by generating the exotic state of odd-frequency superconductivity \cite{Berezinski, Bergeret-2005, Linder-2017}. Odd-frequency Cooper pairs are odd under the exchange of the relative time coordinate between the two electrons forming the pair, in contrast to the conventional equal-time pairing. 
As a consequence, odd-frequency pairing allows the common $s$-wave superconducting state to have spin-triplet symmetry and still satisfy the necessary fermionic nature of superconductivity. In this way, odd-frequency spin-triplet pairing has been evoked to explain the long-ranged superconducting proximity effect measured in superconducting-ferromagnet junctions \cite{Keizer-2006, Bergeret-2001,Bergeret-2005}, but odd-frequency superconductivity has also been found in non-magnetic superconducting junctions \cite{Tanaka-2005}, as well as in bulk multiband SCs \cite{Black-Schaffer-2013} and driven systems \cite{Triola-2017}. 
As equal-time expectation values vanish for odd-frequency superconductivity, it becomes easily a hidden order and direct detection is notoriously hard. Still, odd-frequency superconductivity has been shown to impact physical properties ranging from the Meissner \cite{Abrahams-1995, Yokoyama-2011, Asano-2011, Asano-2015, Bernardo-2015b, Alidoust-2014} and Kerr effects \cite{Komendova-2017,Triola-2018}, to the existence of a finite supercurrent in half-metal (HM) Josephson junctions \cite{Eschrig-2003, Keizer-2006, Asano-2007,Eschrig-2008}. In the last case, the full spin-polarization of the HM completely prohibits spin-singlet superconductivity, but Josephson effect has still been shown to be present in HMs with spin-active interfaces due to the creation of odd-frequency equal-spin triplet pairing \cite{Eschrig-2008}.

\begin{figure}[htb]
\includegraphics[width=\columnwidth]{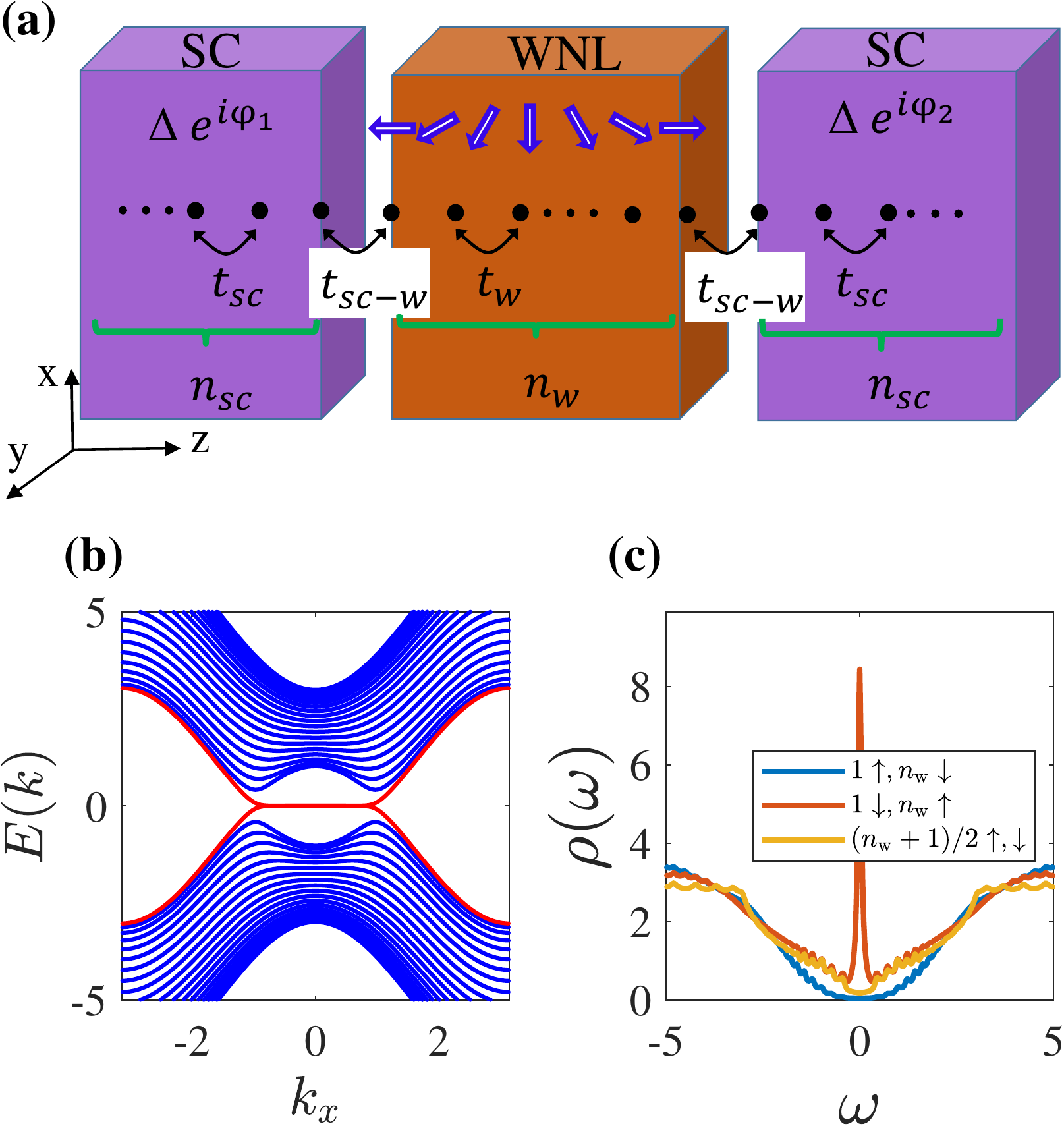}
\caption{\small{{\bf WNL Josephson junction}. (a) Schematic figure of a WNL Josephson junction with the layers in the $z$-direction enumerated in the WNL and all hopping parameters indicated. Arrows indicate schematically how the spin polarization rotates in the WNL. (b) Band dispersion of a $n_\textrm{w} = 21$ layer thick WNL at zero doping $\mu=0$ along the $k_x$ direction, with red indicating the degenerate bands localized at the two WNL surfaces. These midgap states are degenerate since opposite spins behave the same at the two surfaces. (c) Spin-polarized DOS of the first and last layers (red, blue), and middle layer (yellow) of the WNL in (b).}}
\label{fig1}
\end{figure}

In this work, we study a Josephson junction constructed by sandwiching a WNL between two conventional SCs, as shown schematically in Fig.~\ref{fig1}(a). Despite the strong spin-polarization, which naively suppresses any proximity-induced superconductivity and thus Josephson effect, we find a huge Josephson current, even orders of magnitude larger than in normal metal (NM) and HM junctions. We first show how the spin-orbital Weyl interaction in WNLs results in a very efficient creation of equal-spin triplet Cooper pairs, mimicking the spin arrangement in the normal state. It is these equal-spin pairs that carry the Josephson current, which is further dramatically enhanced thanks to the zero-energy drumhead surface states generating excellent interfaces with the SCs. The latter effect of drumhead surface states dramatically enhancing the current we also find in DNL Josephson junctions, but there the Josephson effect is entirely conventional since there is no spin-polarization.
The combined effect of a huge Josephson current and odd-frequency pairing in WNLs creates what can be classified as optimal odd-frequency Josephson junctions, where an experimentally measured Josephson current becomes a direct manifestation of odd-frequency superconductivity. 
%

\section{Results}
\subsection{WNL properties}
We start by establishing the characteristic bulk and surface properties of WNLs.
In order to clearly isolate the most important features we first use a prototype minimal model that captures the important physical details of a WNL. This is a toy model insofar as it captures an accidental nodal loop without any symmetry protection, but its simplicity significantly aids in understanding the underlying physics of WNL Josephson junctions. We then compare these results with a fully generic low-energy Hamiltonian for a WNL derived by the $k\cdot p$ method and show how the two models are in excellent agreement for the Josephson current. 

A minimal WNL toy model is described by the Hamiltonian \cite{Nandkishore-2016, Wang-2017}:
\begin{align}\label{eq:HWLS}
{ H}_\textrm{WNL}=t_\textrm{w}\left(M\sigma_x+2\alpha_2k_z\sigma_y\right)-\mu,
\end{align}
where $M=\alpha_1-k^2$, ${\bf k} = (k_x,k_y,k_z)$ with $k = |{\bf k}|$ is the electron wave vector, and $\boldsymbol{\sigma}$ are the Pauli matrices in spin space.
 The equivalently minimal Hamiltonian for a DNL has instead $\boldsymbol{\sigma}$ in Eq.~\ref{eq:HWLS} acting in orbital basis as a DNL retains spin-degeneracy. Hence the minimal DNL Hamiltonian is described in the same basis as a $4\times 4$ Hamiltonian with two-fold (spin) degeneracy. Moreover, here $t_\textrm{w}$ is the overall hopping amplitude, $\mu$ the chemical potential. For simplicity we measure energy in units of $t_\textrm{w}$.
The Fermi surface is a nodal line loop at zero doping $\mu = 0$, while it forms a thin torus for nonzero $\mu$, with its shape tuned by the two parameters, $\alpha_{1,2}$. We primarily use $\alpha_1 = \alpha_2 =1$, which results in an essentially circular Fermi surface, but our results are not sensitive to this particular choice, see supplementary discussion. The energy dispersion away from zero energy takes the Weyl-like form $k\cdot \sigma$, giving the material class its Weyl nodal loop name \cite{Xie-2015, Sur-2016}. 

We choose to illustrate the physics of WNL Josephson junctions using the minimal Hamiltonian in  Eq.~\ref{eq:HWLS} as it is a very simple model, offer easy comparison between WNLs and DNLs, and is also used in earlier literature discussing the immunity of the surface of WNLs towards conventional superconductivity \cite{Wang-2017}. Still, Eq.~\ref{eq:HWLS} only uses two out of the three $\sigma$-matrices, while in a generic WNL all three $\sigma$-matrices will be represented.
Thus, in addition to Eq.~\ref{eq:HWLS}, we present complementary results for a generic low-energy Hamiltonian, which includes all three $\sigma$-matrices:
\begin{align}\label{eq:Hamil-g}
{ H}_\textrm{g-WNL}=t_\textrm{w}\left(M\sigma_x+2\alpha_2k_z(k_x^2-k_y^2)\sigma_y+4\alpha_2k_xk_yk_z\sigma_z\right)-\mu,
\end{align}
which is the same low-energy Hamiltonian as proposed for the WNL HgCr$_2$Se$_4$ \cite{PRL.107.186806, PRL.121.166802} upon a 90$^\circ$ rotation of the spin axes.
The generic Hamiltonian Eq.~\ref{eq:Hamil-g} breaks the ${\cal T}{\cal I}$ symmetry, i.e.~the product of time-reversal ${\cal T}=-\textrm{i}\sigma_y{\cal K}$, where ${\cal K}$ is the complex conjugation operator, and spatial inversion ${\cal I} : (x,y,z)\rightarrow (-x,-y,-z)$ symmetries, allowing for finite spin-polarization along with spin-orbit coupling. It also contain an additional mirror symmetry,$ {\cal M}:(x,y,z)\rightarrow (x,y,-z)$, to protect the nodal line \cite{PRL.107.186806}.

To study a finite WNL with its surface states, we place the continuum Hamiltonians Eqs.~\eqref{eq:HWLS}-\eqref{eq:Hamil-g} on cubic lattice, performing the usual substitution of $k \rightarrow \sin(k)$ and $k^2 \rightarrow 2(1-\cos(k))$. In the $x$- and $y$-directions we keep the reciprocal space description (implemented by applying periodic boundary conditions), while in the $z$ direction we discretize the Hamiltonians to produce a finite slab, see Methods. 
We assume a lattice with $n_\textrm{w} =21$ layers along the $z$-direction, as indicated by black solid circles in Fig.~\ref{fig1}(a). Note that the distance between lattice sites are that of the full unit cell and therefore $n_\textrm{w}=21$ junction length is reasonable for the Josephson effect, but we have also checked our results for much longer lengths, see supplementary discussion.

In order to understand the basic physics of minimal WNLs, we plot in Fig.~\ref{fig1}(b) the energy dispersion along the $k_x$ direction for the Hamiltonian in Eq.~\eqref{eq:HWLS}. At zero doping the band dispersion is electron-hole symmetric and two bands, indicated in red, have a vanishing electron group velocity and form zero-energy flat bands in a large region around the $\Gamma$ point. 
These two bands reside on the two slab surfaces, as clearly seen in Fig.~\ref{fig1}(c), where we display the spin-polarized DOS for several different layers. The two surface layers have a very large peak at zero energy, which are fully spin-polarized but in opposite directions.
The bulk on the other hand has only a small constant DOS at zero energy due to the nodal line Fermi surface and there is no net spin-polarization in the middle of the slab. 
It is the spin-orbit-like interaction term $\alpha_2\sigma_y\sin(k_z)$ in the WNL Hamiltonian Eq.~\eqref{eq:HWLS} that causes the characteristic spin rotation throughout a WNL material. This term is off-diagonal in spin-space and hence couples the up and down spins in a spatially dependent manner. In a slab configuration, this term rotates the spins from spin-down polarization at zero-energy for the left surface (layer 1) to spin-up polarization for the right surface (layer $n_\textrm{w}$), which we  schematically illustrate with arrows in Fig.~\ref{fig1}(a). It is only at very large energies that the first ($n_\textrm{w}$th) surface hosts any up- (down-)spin polarization. At finite doping the drumhead surface states remain unchanged with a full spin-polarization, but are now located at an energy $\mu$ below the Fermi level. At the same time, the DOS at the Fermi level in the bulk increases due to the torus-shaped Fermi surface at finite doping. The spin-polarization stays however almost complete in a rather large energy window around $\mu =0$ and thus results are not sensitive to the exact tuning of the chemical potential.

\subsection{Odd-frequency pairing}

Next, we place two conventional spin-singlet $s$-wave SCs of the same superconducting material in proximity to the two surfaces of the WNL slab, see Fig.~\ref{fig1}(a). The superconducting order parameter amplitude in the SCs is set by $\Delta$, but we allow for different phases, $\varphi_\textrm{L,R}$ such that a Josephson current can be generated across the superconducting heterostructure. We couple the WNL and the SCs using a generic spin-independent tunneling amplitude $t_\textrm{sc-w}\sigma_0$, which connects the surface site of the left and right SCs to the left and right surfaces of the WNL, respectively.

To study proximity-induced superconductivity in the WNL we extract the superconducting pair amplitudes in the WNL by calculating the anomalous Green's function $F$ of the full heterostructure (see Methods).
Here we only have to focus on isotropic, or equivalently on-site, $s$-wave pairing as we find that to be the only relevant spatial symmetry of all pair amplitudes. Notably, all the $p$-wave components in the $xy$-plane are practically zero for both WNL models; Eq.~\eqref{eq:HWLS} even hosts an explicit in-plane spatial even parity preventing in-plane odd-parity proximity pairing.
In the supplementary discussion we additionally show that out-of-plane $p$-wave pairing is also much smaller, stemming from it not being aligned with the superconducting surface, and also that extended $s$-wave pairing is not relevant.
Moreover, we note here that, if we were to add any disorder, that would even further favor isotropic $s$-wave pairing, thus only strengthening our results.
In terms of spin configurations, the Weyl spectrum rotates the spin and can thus allow for both equal- and mixed-spin triplet pairs. We therefore study all possible spin configurations for the superconducting pairing.

%

\begin{figure*}[htb]
\includegraphics[width=\linewidth]{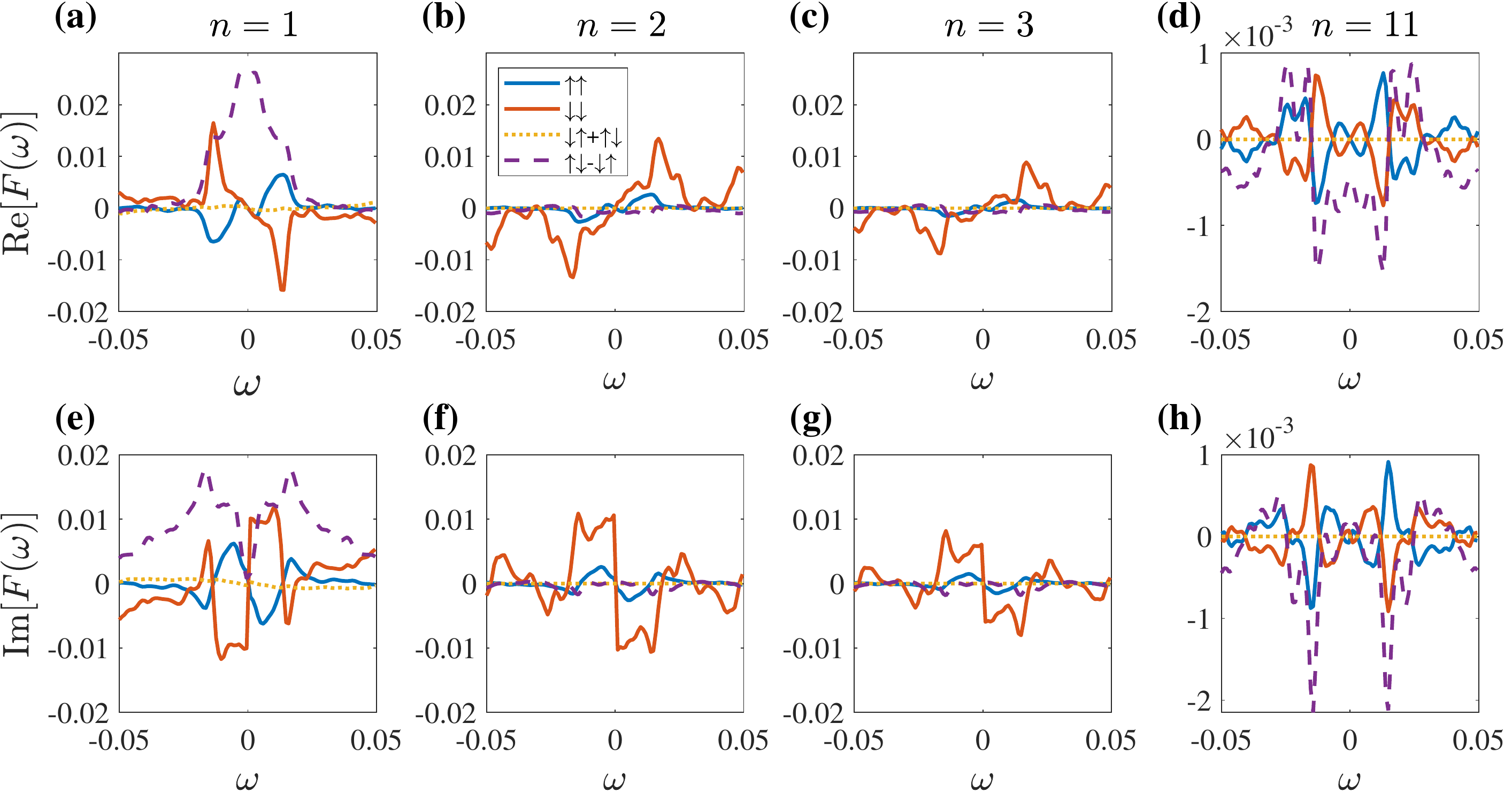}
\caption{{\bf Pair amplitudes in WNL Josephson junctions and their frequency dependence.} Real (upper panels, a-d) and imaginary (bottom panels, e-h) parts of the anomalous Green's function $F$ as a function of frequency $\omega$, capturing the pair amplitudes divided into the equal-spin ($\uparrow\uparrow$ and $\downarrow\downarrow$), mixed-spin ($\uparrow\downarrow+\downarrow\uparrow$) triplet, and spin-singlet ($\uparrow\downarrow-\downarrow\uparrow$) components. Left to right panels shows results for the $n=1,2,3$ and middle $n=(n_\textrm{w}+1)/2=11$ layers, respectively. 
Here we use Eq.~\eqref{eq:HWLS} with $\mu_\textrm{w}=0$, $t_\textrm{sc-w}=0.5$, $\Delta=0.01$, $t_\textrm{sc}=1$, $\mu_\textrm{sc}=2$, and $\varphi_\textrm{L}=\varphi_\textrm{R}=0$, .}
\label{fig:Fw}
\end{figure*}

In Fig.~\ref{fig:Fw} we display the real (upper panels, a-d) and imaginary (lower panels, e-h) parts of the anomalous Green's function $F$ as a function of frequency $\omega$, divided into all possible spin configurations for the minimal WNL model. Each column represent a different layer in the WNL, layers $1,2,3$, and $n=(n_\textrm{w}+1)/2$. We note directly that all different spin-triplet components appear throughout the WNL and that they are always odd functions of frequency, as required by the Fermi-Dirac statistics of the Cooper pairs.
In the first surface layer there is still notable spin-singlet pairing. This is to be expected since the first layer is directly coupled to the SC and therefore necessarily harbors superconducting pairs of same symmetry as in the SC. However, the spin-singlet amplitude decay extremely quickly into the WNL, such that it has essentially disappeared already in the second layer. 
This behavior is not surprising when considering that the drumhead states of the WNL are fully spin-polarized and thus the tunneling of opposite spins is energetically extremely costly. Similar immediate destruction of spin-singlet amplitudes have previously also been reported for HM junctions \cite{Asano-2007}. 
Despite the complete lack of proximity effect for spin-singlet superconductivity beyond the first surface layer, there is still significant pairing induced in the WNL. It is instead equal-spin triplet pairing with spins aligned with the spin-polarization of the drumhead state that growths and heavily dominates in the subsurface layers. Thus, the WNL essentially becomes an odd-frequency superconductor beyond the very first surface layer.

In the middle of the sample, Fig.~\ref{fig:Fw}(d,h), all pair amplitudes are suppressed due to the distance from the SC interface, but notably, the two equal spin pairing terms have exactly the same magnitude, just mirrored in $\omega = 0$. Plotting the pair amplitudes also for the right half of the WNL, we find exactly the same results as for the left part shown in Fig.~\ref{fig:Fw}, only with spin-up and spin-down interchanged. 
The behavior of the equal-spin pairing is the superconducting equivalent of the spin-polarization in the normal state twisting from full spin-down polarization in the left surface layer to full spin-up polarization in the right surface layer. 
Thus the appearance of large odd-frequency equal-spin triplet components in WNLs is guaranteed by the intrinsic Weyl spin-orbital structure of the WNL normal state.

%
\begin{figure*}[htb]
\includegraphics[width=0.75\linewidth]{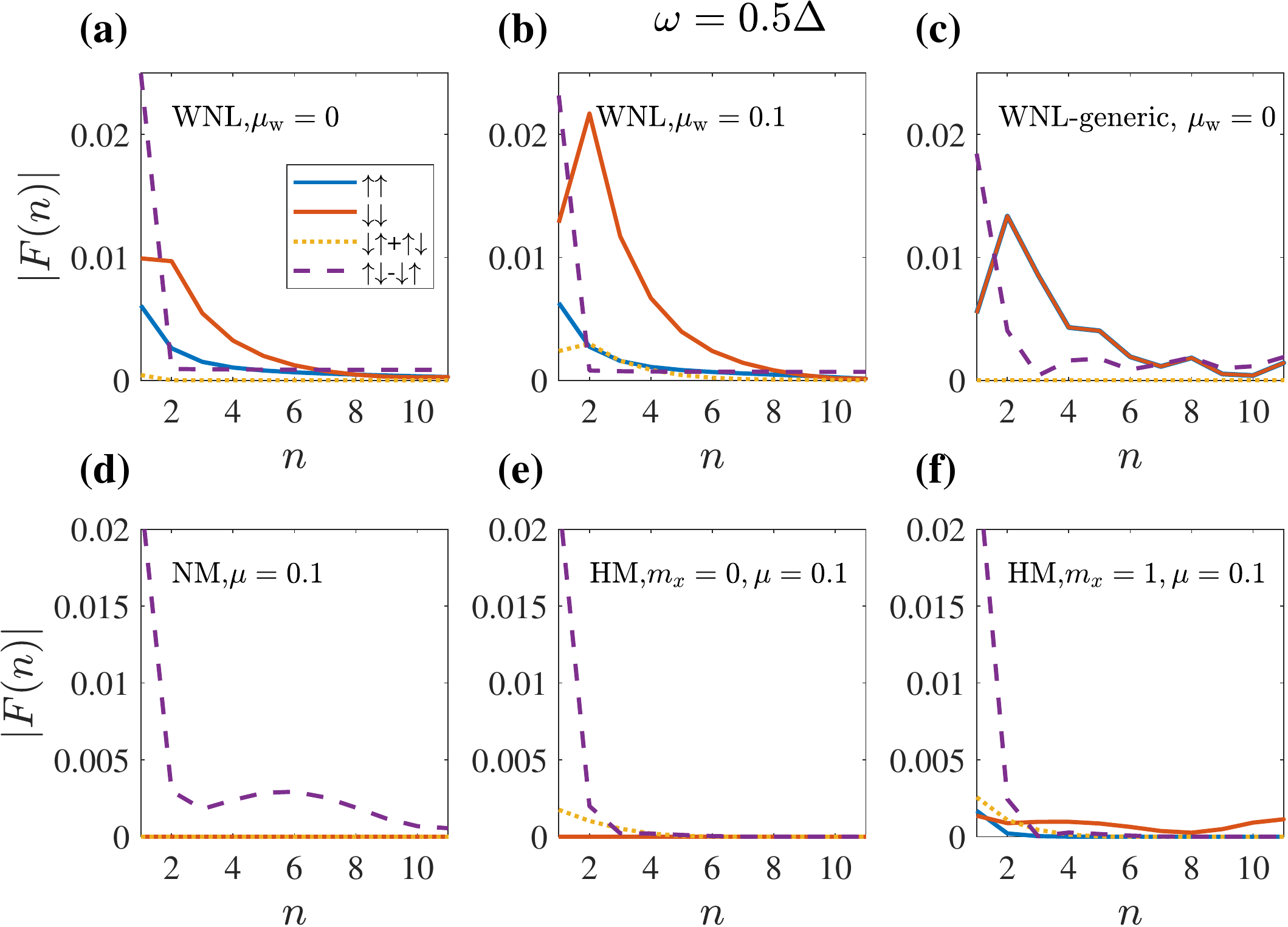}
\caption{{\bf Pair amplitude decay in WNL Josephson junctions compared to NM and HM junctions.} Evolution of the absolute value of equal-spin ($\uparrow\uparrow$ and $\downarrow\downarrow$), mixed-spin ($\uparrow\downarrow+\downarrow\uparrow$) triplet, and  spin-singlet ($\uparrow\downarrow-\downarrow\uparrow$) pair amplitudes into the middle of the WNL within the (a,b) minimal model Eq. \eqref{eq:HWLS} with chemical potential set to $\mu_\textrm{w}=0,0.1$, respectively, and (c)generic model Eq. \eqref{eq:Hamil-g} with $\mu_\textrm{w}=0$. It is compared to similar junctions with NM (d) and HM without (e) and with (f) an spin-active interface. Pair amplitudes are extracted for $\omega = 0.5\Delta$.
Same parameters as in Fig.~\ref{fig:Fw} for the WNL, while the NM and HM are the spin-independent part of the WNL with the HM having an additional $m_z\sigma_z$ with $m_z =0.5$ magnetization in the bulk and an interface $m_x\sigma_z$ with $m_x=1$ term modeling the spin-active interface.
}
\label{fig:Fs}
\end{figure*}

To better probe the propagation of Cooper pairs inside the WNL, we plot in Figs.~\ref{fig:Fs}(a,b) the absolute value of the different pair amplitudes for the minimal WNL in  Eq.~\eqref{eq:HWLS}, as a function of the layer index $n$ for the whole left side of the WNL. 
The pair amplitudes in the right part is obtained from the left layers by just interchanging spin-up and spin-down. We display the result for two different chemical potentials, $\mu=0,0.1$, respectively, to capture both nodal loop and torus-shaped Fermi surface WNLs. Further, we set $\omega=0.5\Delta$, but keep all other parameters as in Fig.~\ref{fig:Fw}. Other choices of $\omega$ can be found in the supplementary discussion, showing no change of trends compared to Fig.~\ref{fig:Fs}. 
We find the same extremely fast suppression of the spin-singlet amplitude for both the undoped and doped case. The mixed-spin triplet state experiences the same decay, due to the same unfavorable spin alignment as the spin-singlet pairing. 
Instead, it is spin-down triplet pairing that is clearly dominating, also well into the WNL and for all $\mu$.
The finite but still small spin-up triplet pair amplitude is due to probing the pair amplitudes at finite energies, where also bulk states give a finite contribution. 
Increasing the doping level thus show no significant changes in the relative importance of the different pairing channels. However, the magnitude of the pair amplitudes increases due to the increased bulk DOS in finite doped WNL. Overall this shows that the almost exclusively odd-frequency pairing state in the WNL is not sensitive to the tuning of the doping level. The relative insensitivity to the doping level means no doping fine-tuning is needed nor can the effects be fragile to a finite buckling of the drumhead surface state. In the supplementary discussion we also plot the pair amplitude propagation for other lengths of the WNL junction and confirm that dominating odd-frequency pairing is also present for much longer junctions. Thus the presence of odd-frequency pairing is not just a simple surface effect, but more appropriately linked to the Weyl spin-orbital structure of the bulk band structure.

Furthermore, in Fig.~\ref{fig:Fs}(c) we present the evolution of pair amplitudes based on a generic WNL given in  Eq.~\eqref{eq:Hamil-g}, also at $\mu =0$ to compare with panel (a). As seen, the spin-singlet pairing decays very fast, although it still have a finite amplitude in the middle of the sample. This is due to a lack of full spin-polarization on the surfaces. Thus the spin-singlet response is rather similar to that of the slightly doped minimal WNL in panel (b). 
Dominating inside the WNL are clearly the  equal-spin triplet states, and they are even stronger in comparison to the equal-spin triplet pairing in the minimal WNL in panel (a), which we can attribute to more strongly spin-splitted bands in the generic WNL. At the same time, the bands in the generic WNL are symmetric under interchange of spin and changing $k \rightarrow -k$, and thus both equal-spin triplet states after summation over $k$ are equal.

To demonstrate the remarkable pairing in WNL Josephson junctions we compare the results with the behavior of similar Josephson junctions made of NM (d) and HMs without (e) and with (f) a spin-active interface region. In order to create systems with directly comparable properties we model the NM by removing all spin-dependence from the WNL Hamiltonian in Eq.~\eqref{eq:HWLS}, such that $H_\textrm{NM}=t_\textrm{w}(6-2\cos(k_x)-2\cos(k_y)-2\cos(k_z))-\mu$. This creates a prototype parabolically dispersive metal, where we set $\mu=0.1$ to reach a finite bulk DOS. For the NM, no term breaks the spin degeneracy, and thus only spin-singlet pairing is present in the NM. This spin-singlet amplitude experiences a regular slow decay, set by the conventional proximity effect.

For the HM junctions, we add the term $m_z\sigma_z$, with $m_z=0.5$, to the NM Hamiltonian $H_\textrm{NM}$, such that only spin-down electrons are present at zero energy. This strong magnetization causes the same dramatic suppression of the spin-singlet amplitude as in the WNL. It also allows for spin-rotation into the mixed-spin triplet state. However, this odd-frequency spin-triplet pairing state is always small and fast-decaying, as the occurrence of spin-up components is not energetically favored by the magnetization \cite{Parhizgar-2014}. 
To achieve a spin-down triplet component, an additional spin quantization axis has to be present in the HM junction. This is often achieved by introducing a spin-active region at the interface between the SC and HM, see e.g.~\cite{Asano-2007, Eschrig-2008}. In this case the Cooper pair spin quantization axis rotates between the interface and the HM bulk, with the consequence that spin-equal pairing is induced beyond the interface. In Fig.~\ref{fig:Fs}(f) we therefore add a term $m_x\sigma_x$, with  $m_x=1$, to the the two surface layers of the HM to model a strongly spin-active interface. 
As a result, spin-equal triplet pairing is generated, of which the spin-down component survives throughout the HM region. The spin-up triplet state is also initially generated at the interface, but it is energetically unfavorable and decays very quickly in the HM. 

Comparing the WNLs with the NM and HM junctions, we see that the WNLs junction closest resemble that of the HM with an active spin interface, since they both experience a strong proximity effect consisting of odd-frequency equal-spin triplet pairing. However, WNL Josephson junctions are fundamentally different from HM Josephson junctions as they do not need any additional spin-active interface region added during manufacturing in order to generate equal-spin odd-frequency pairing.
In the simplistic WNL junctions based on Eq.~\eqref{eq:HWLS} we find that the intrinsic Weyl spin-orbit coupling causes the initial spin-down pairing in the left surface to rotate into spin-up pairing in the right surface. As a consequence, there is a clear decay of the spin-up triplet component into the middle of the WNL, which is not present in the HM. It is thus not the distance from the SC that causes the main decay of the equal-spin triplet components in Fig.~\ref{fig:Fs}(a,b), but mainly the continuous rotation of the spin orientation of the Cooper pairs. This is also clear when considering other WNL junctions lengths in the supplementary discussion. For the generic WNL model, we find that both equal-spin triplet pairing are present at equal amounts in each layer due to its spin structure, and then there is a smaller decay of the odd-frequency state away from the surface of the WNL.

We also note that the size of the pair amplitudes in the WNLs at zero doping is actually of the same order of magnitude as in the NM junctions. This is particularly surprising since the nodal line bulk state has a much smaller DOS at low energies compared to the NM (see supplementary discussion for a detailed account on the low-energy DOS). 
We attribute the large pair amplitudes in undoped WNLs to the singular zero-energy DOS of the drumhead surface states, which creates a naturally strong coupling between the WNL and SCs. The large effect of the surface drumhead states is further evident when we compare the results for a DNL Josephson junction. Here only spin-singlet $s$-wave pairing is present since the DNL Hamiltonian is spin-degenerate, but we find a very large conventional proximity effect due to the the (un-polarized) surface drumhead states, see supplementary discussion for details.
%


\begin{figure*}[htb]
\includegraphics[width=0.75\linewidth]{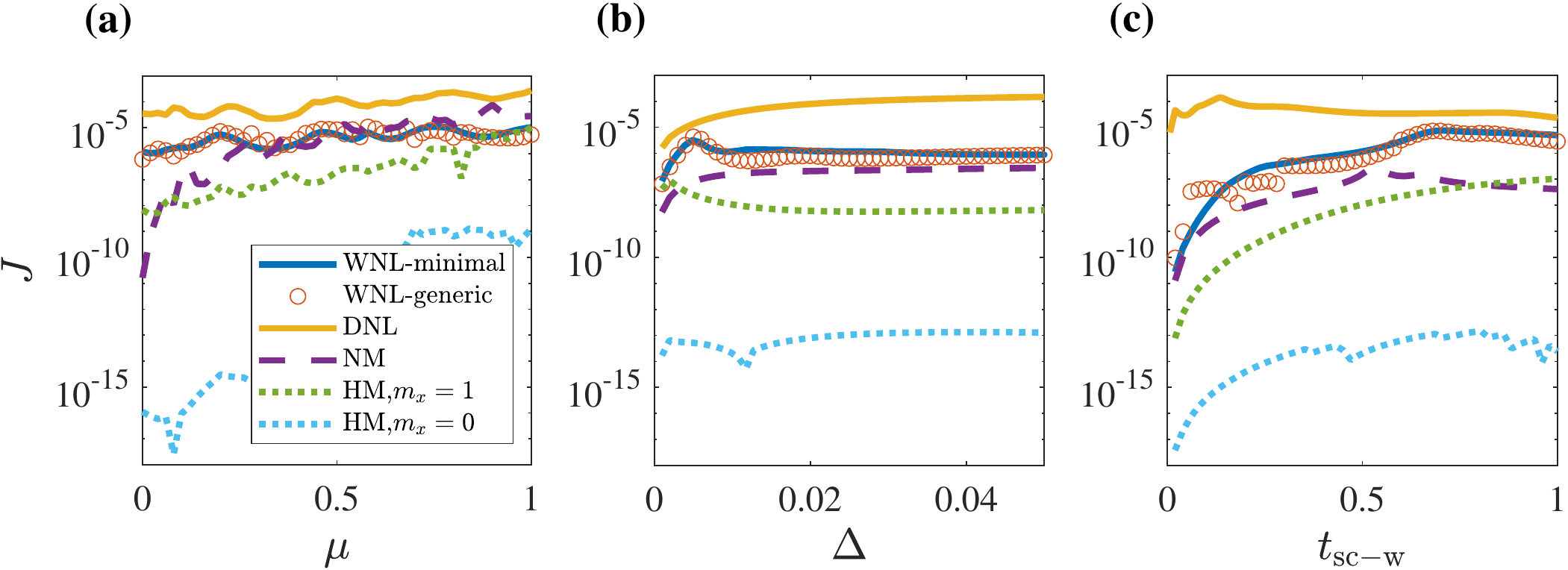}
\caption{{\bf Josephson current in WNL Josephson junctions compared to DNL, NM, and HM junctions.} Parameter dependencies of the Josephson current (in units of $\textrm{e}a^2t_\textrm{w}\hbar^{-1}$) in WNL, DNL, NM, and HM Josephson junctions as a function of chemical potential $\mu$ (a), order parameter amplitude $\Delta$ (b), and tunneling between the SC and junction material $t_\textrm{sc-w}$ (c). Fixed parameters are the same as in Figs.~\ref{fig:Fw} and \ref{fig:Fs} except $\varphi_\textrm{L}={\rm \pi}/2,\varphi_\textrm{R}=0$ and in (b,c) NM and HM have $\mu=0.1$, while $\mu_\textrm{w} =0$ in the WNL. }
\label{fig.Jmu}
\end{figure*}

\subsection{Exotic Josephson current}
Having shown how odd-frequency spin-triplet pairing generates large proximity-induced superconductivity in WNLs, despite the incompatibility of spin-polarization and the spin-singlet SCs, we now turn to the possibility of measuring a finite Josephson current in the WNL junctions. For comparison we also calculate the current in DNL, NM, and HM Josephson junctions. While we limited the investigation in Figs~\ref{fig:Fw}-\ref{fig:Fs} to isotropic $s$-wave pairing, the contributions from all pairing channels are automatically included when calculating the current, although all non-$s$-wave contributions are small for WNLs. 

In Fig.~\ref{fig.Jmu} we show in a log-plot the maximum Josephson current $J$ between two conventional spin-singlet SCs as a function of the chemical potential $\mu$, superconducting order parameter $\Delta$, and WNL-SC tunneling $t_{\rm sc-w}$. The results are obtained for $\varphi_\textrm{L}={\rm \pi}/2,\varphi_\textrm{R}=0$, which to a very good approximation gives the maximum current as $J \sim \sin(\varphi_\textrm{L}-\varphi_\textrm{R})$, see supplementary discussion. The most remarkable result is that WNL junctions carry a large Josephson current through all parameter regimes, orders of magnitude higher than the HM junctions, the other odd-frequency systems, but also compared to the NM junctions for a large range of doping levels. The current for both WNL junctions is also always essentially identical, illustrating explicitly that the specific form of the WNL Hamiltonian is irrelevant for the Josephson current. We note particularly that the slightly different spin-singlet pairing in the two different WNL junctions do not influence the Josephson effect. This can also be understood from the results for NM junction, which only contains spin-singlet pairing but where the Josephson current is orders of magnitude smaller than the WNL junctions in the small doping regime.
Instead it is the common physical properties of WNLs, their strong spin-polarization (surface and bulk) and drumhead surface states, that generates a unique odd-frequency system with large Josephson currents. WNL Josephson effect thus forms a key example on the importance of odd-frequency pairing in inhomogeneous superconducting systems. If it were not for the odd-frequency correlations, there would be no proximity-induced superconductivity or measurable Josephson effect in WNL junctions.

In particular, we find for increasing $\mu$ in Fig.~\ref{fig.Jmu}(a) that the current increases for all junctions, while also displaying some overlaid oscillatory behavior. This is expected since the low-energy DOS increases with $\mu$ for all junctions with some smaller oscillations due to finite size effects. Most notably, the WNL junctions carry a very large current over a wide range of low to moderate doping levels, even orders of magnitude larger than both the NM and HM junctions. This is really quite exceptional considering the low bulk DOS of the nodal line/thin torus Fermi surface in the bulk of WNLs. In fact, the WNL bulk DOS for Eq.~\eqref{eq:HWLS} is for the full doping range smaller than that of the bulk NM, for $\mu\lesssim 0.4$ it is even several times smaller, see supplementary discussion. We must therefore accredit the large Josephson current in WNL junctions to a remarkable effect of the drumhead surface states.

The effect of the drumhead states is even more obvious when we compare to a DNL Josephson junction. Here we use a DNL with the same band structure and total DOS as the WNL in Eq.~\eqref{eq:HWLS}, but keep a fully spin-degenerate ground state. Thus the Josephson effect and current are entirely conventional and carried by even-frequency spin-singlet $s$-wave pairs in a DNL, while the coupling to the SCs is determined by the same surface DOS peak as in the WNL junction. We find that the DNL junction carries a current that is consistently large over all parameter regimes investigated in Fig.~\ref{fig.Jmu}. In the simplest of cases, the maximum Josephson current in a junction is proportional to the normal current, which in turn is proportional to ${\rm e}^2 \rho v_\textrm{F}$, where $\rho$ is the DOS and $v_\textrm{F}$ is the Fermi velocity in the direction of the current. While the DOS varies within the junction for both DNLs and WNLs, this still demonstrates that the drumhead surface state should dramatically enhance the current in both cases. 
We here note that in general flat band systems the current usually suffers from the diminishing Fermi velocity. However, for the DNL and WNL junctions considered here it is only the Fermi velocity in-plane that is quenched due to the flat dispersion, while it is the velocity in the out-of-plane direction that determine the current, which is still large. 
We also note that the DNL current is larger than the WNL current, which we can easily attribute to the DNL junction only hosting conventional pairing and thus does not have to transform to other pair amplitude symmetries to create a viable Josephson effect. 

Moving on to compare with the HM junctions that also host odd-frequency spin-triplet pairs, we find that the HM without a spin-active interface cannot effectively carry Josephson current, which is a direct consequence of the lack superconducting pair amplitudes in the HM. Introducing a spin-active interface layer, we find a significantly increased current due to the perseverance of odd-frequency equal-spin triplet pairs inside the HM. Yet, it is only at extremely large $\mu$ that the HM system carries a similarly sized current to the WNL junction. For large $\mu$ the drumhead surface states are located very far from the Fermi level and are thus less active in electric transport. Thus, the large current in heavily doped WNL junctions is instead primarily a manifestation of how powerful the Weyl spin-orbit interaction is in generating odd-frequency equal-spin pairing to carry the current, and thus it is natural that the WNL and HM junction behave similarly in this extreme doping limit.

We here note that enhancement of Josephson current in junctions with odd-frequency pairing has previously been associated with generation of zero-energy Andreev bound states, as for example in SC-insulator-SC junctions \cite{Tanaka-1996, Barash-1996, Tanaka-1997, Kwon-2004}. In fact, zero-energy Andreev bound states have been found to be very common in systems with odd-frequency pairing \cite{Tanaka-2007, Tanaka-2007-2, Tanaka-2012}, although not always present \cite{RobinsonLinder, Black-SchafferBalatskyTI, Black-SchafferBalatskyTI2, Black-SchafferBalatskyMultiband}. In the supplementary discussion we show that zero-energy Andreev boundary states also exists at the WNL-SC interface. However, considering the much larger current for the WNL junction as compared to the HM junctions, zero-energy Andreev bound states alone cannot alone explain the results, but instead the drumhead surface states are more important.

Finally, turning Figs.~\ref{fig.Jmu}(b,c), we explore the other parameter dependencies of the Josephson current. Again note that both models of WNL are in a excellent agreement with each other, and thus our main conclusions are not dependent on any specific model.
Fig.~\ref{fig.Jmu}(b) shows the variation of the Josephson current with respect to the superconducting gap parameter $\Delta$, keeping the chemical potential of NM and HM at $\mu=0.1$, while the WNLs are both undoped. Thus, both the NM and HM have a large metallic DOS at low energies, while the DNL and WNL have only 1D Fermi nodal loops. Despite this we see how the Josephson current is actually larger in WNL junctions compared to the NM and HM junctions for all $\Delta$ values. Notably, the choice of $\Delta = 0.01$ in Figs.~\ref{fig:Fw} and \ref{fig:Fs} gives by no means the maximum Josephson current, for example choosing $\Delta\approx 0.005$ gives approximately three times larger current.
In Fig.~\ref{fig.Jmu}(c) we then tune the tunneling between the SCs and the junction, $t_\textrm{sc-w}$. 
Usually this parameter is lower than the hopping inside the junction and in the SCs, limiting it to $t_\textrm{sc-w}<1$, and in practice it is tuned by modifying the interfaces. 
Larger tunneling clearly enhances the Josephson current in all junctions, but we find that the WNL junctions again carry larger Josephson current than the NM and HM junctions for all values of $t_\textrm{sc-w}$. 
In particular, a large Josephson current is present even for small $t_\textrm{sc-w} \sim 0.1$, where the overall effect of the SCs on the electronic structure of the WNL is necessarily quite limited.
We provide additional plots for other choices of chemical potential in NM and HM in the supplementary discussion, verifying that the large Josephson currents through the WNL junctions in Fig.~\ref{fig.Jmu} are generic results and not restricted to a narrow range of physical parameters.

\section{Discussion}
To summarize we establish a large and exotic Josephson effect in WNL superconducting junctions, driven by spin-polarized surface and bulk states and bulk Weyl spin-orbit interaction. Combined these effects enable a very effective transformation of the original spin-singlet pairing in the SC leads into an odd-frequency equal-spin triplet pairing state, which then carries the Josephson current into the WNL. Further, the singular DOS of the drumhead surface states make carrier transport between the SCs and the WNL very effective, which significantly enhances the current. In fact, we find that the WNL Josephson current can easily be orders of magnitude larger than the current in NM Josephson junctions, despite a much lower bulk DOS in the WNL due to its nodal line Fermi surface. A similar huge Josephson current also exists for DNL due to its equivalently large surface DOS peak, but here the current is carried by conventional spin-singlet superconductivity and is thus an expected result. 
Notably we find exceptionally similar Josephson currents in both simple and generic models for WNLs. Thus our results are not sensitive to the particular choice of model for the WNL but depend on the general features of WNLs.

The physics of the WNL Josephson junction can be understood from the behavior of HM junctions with spin-active interfaces, as both junctions have dominating odd-frequency equal-spin triplet pairing carrying the Josephson current. However, the strong bulk spin-orbit coupling in the WNL Josephson junction make it more similar to HM junctions with a helical magnet configuration instead of just a spin-active interface. Such helical magnets have recently become the prototype experimental odd-frequency system due to their large and dominating odd-frequency response probed by both Josephson effect and paramagnetic Meissner effect \cite{Robinson-2010, Linder-2015, Bernardo-2015a, Bernardo-2015b}. 
Still, the WNL is a much more optimal odd-frequency Josephson link due to the drumhead surface states allowing for excellent interfacial coupling and thus dramatically larger currents. In fact, this huge Josephson current would not even have existed if it was not for odd-frequency superconductivity.
The importance of the drumhead surface states persists as long as the junction is not much longer than the decay length of the conventional superconducting proximity effect, as then most pairing occurring in the surface states still propagate through the junction. Thus the WNL odd-frequency Josephson current exists on the same length scales as conventional Josephson effect.
In conclusion, the combination of finite spin-polarization, Weyl spin-orbit interaction, and drumhead surface states makes WNLs optimal odd-frequency materials, with the odd-frequency pairing detectable by finite and unexpectedly large Josephson currents.

\small
\section{Methods} 
For the WNL we consider a finite slab in the $z$-direction. We use the reciprocal-space continuum Hamiltonian in Eq.~\eqref{eq:HWLS} for $(k_x,k_y)={\bf k}_\parallel$, while discretizing the model in the $z$-direction in the standard way: $t_\textrm{w}\cos(k_z)$ becomes the nearest neighbor hopping $t_\textrm{w}$, while $t_\textrm{w} \sigma_y \sin(k_z)$ generates a nearest neighbor spin-orbit interaction. Moreover, in the $x,y$-directions  we perform the substitution $k \rightarrow \sin(k)$ and $k^2 \rightarrow 2(1-\cos(k))$, to comply with a lattice description. For concreteness we use $n_\textrm{w} = 21$ layers in the $z$-direction, but our results are not sensitive to the number of layers, granted that the surface states are spatially well separated and the junction still carry a supercurrent for a NM. 
By calculating the retarded Green's function $G=(\omega+\textrm{i}0^+-{H})^{-1}$ we obtain the spin-resolved DOS in each WNL layer $n$ as $\rho_{n,\sigma}=-\frac{1}{{\rm \pi}} \Im \sum_{{\bf k}_\parallel}G_{n,\sigma}({{\bf k}_\parallel,\omega+\textrm{i}0^+})$.

To investigate Josephson junctions we attach a conventional SC to each of the two WNL slab surfaces with the Hamiltonian in Nambu space being
\begin{align}\label{HSC}
{\cal H}_{sc}^j=\begin{pmatrix} h_\textrm{sc}\sigma_0 & \Delta \textrm{e}^{\textrm{i}\varphi_j}\sigma_y\\ \Delta \textrm{e}^{-\textrm{i}\varphi_j}\sigma_y & -h_\textrm{sc}\sigma_0
\end{pmatrix},
\end{align}
where $j=\textrm{L,R}$ indicates the left and right SC, respectively. We define the SCs on a simple cubic lattice, where the normal state Hamiltonian in $(k_x,k_y)$ takes the form $h_\textrm{sc}=t_\textrm{sc}(2-\cos(k_xa)-\cos(k_ya)) -\mu_\textrm{sc}$ due to translational invariance, while in the $z$-direction we have the nearest neighbor hopping $t_\textrm{sc}$ between the different layers of the SCs. Spin-singlet $s$-wave superconductivity is implemented as usual by an on-site order parameter $\Delta \textrm{e}^{\textrm{i}\varphi_j}$.
We use $n_\textrm{sc}=20$ layers for each SC and set $t_\textrm{sc} = 1$ and $\mu_\textrm{sc} = 2$ to create SCs with large low-energy DOS. The WNL and SCs are connected by a generic spin-independent tunneling $t_\textrm{sc-w}$. The results are not qualitatively sensitive to $n_\textrm{sc}$ or other physical parameters in the SCs, as shown in the supplementary discussion. 
As a result, the total Hamiltonian takes the form
\begin{align}
{\cal H}_t=\begin{pmatrix}
{\cal H}_\textrm{SC}^\textrm{L} & T_\textrm{L} & 0\\
T_\textrm{L}^\dagger &{\cal H}_\textrm{W} & T_\textrm{R}^\dagger\\
0&T_\textrm{R} & {\cal H}_\textrm{SC}^\textrm{R}
\end{pmatrix},
\end{align}
where ${\cal H}_\textrm{W}$ is the Nambu representation of the WNL Hamiltonian in Eqs.~\eqref{eq:HWLS} or \eqref{eq:Hamil-g}. Moreover, the matrices $T_\textrm{L/R}$ are $4n_\textrm{sc} \times 4 n_\textrm{w}$ matrices that connects the corresponding surfaces of the WNL to the left and right SCs.

Using the same Green's function technique but now the whole system WNL + SCs, we extract the anomalous electron-hole part, which is proportional to $\langle c_{n\sigma}^\dagger c_{n\sigma'}^\dagger\rangle$, with $c_{n\sigma}^\dagger$ the electron creation operator in layer $n$ with spin $\sigma$, and thus gives the pair correlations in all SC and WNL layers. By integrating over $\textbf{k}_\parallel$ we obtain the $s$-wave contribution, which we report individually for each pair spin configuration. 
 Note that in order to achieve the correct pair amplitudes as a function of $\omega$, we need to use the advanced (retarded) Green's function for negative (positive) frequencies \cite{Linder-2017,Cayao-2017}.
 
To calculate the Josephson current between two SCs, we use the continuity equation ${\boldsymbol \nabla}\cdot {\bf J}+\langle\partial \hat{\rho}_n/\partial t\rangle =0$, where ${\bf J}$ is the current density vector and $\hat{\rho}_n=\sum_\sigma c^\dagger_{n\sigma} c_{n\sigma}$ is the density operator \cite{Black-Schaffer-2008, Bjornson-2015}. Since we are working only with layers in the $z$-direction, we sum over all ${\bf k}_\parallel$. Moreover, $\langle ... \rangle$ indicates the expectation value taken over whole system, which we obtain by summing over all occupied energy eigenstates. Finally, the derivative of the density operator $\hat{\rho}_n$ can be obtained from $\langle\partial \hat{\rho}_n / \partial t\rangle=\langle [H,\hat{\rho}_n]\rangle$. 
Here the right-hand side generates terms of the type, $c^\dagger_nc_{n+1}$ and $c^\dagger_nc_{n-1}$, which are intuitively proportional to the in- and out-going currents ($j_\textrm{in},j_\textrm{out}$) in each layer. Writing ${\boldsymbol \nabla}\cdot {\bf J}=(j_\textrm{out}-j_\textrm{in})/a^2$, we can obtain the Josephson current $J=\textrm{e}(j_\textrm{out}-j_\textrm{in})$ in units of $\textrm{e}a^2t_\textrm{w}\hbar^{-1}$.  

\normalsize
\section{Acknowledgment}
We thank A.~Balatsky, K.~Bj\"ornson, J.~Cayao, P.~Dutta, S.~Nahas, and C.~Triola for fruitful discussions and acknowledge financial support from the Swedish Research Council (Vetenskapsr\aa det, Grant No.~621-2014-3721), the G\"{o}ran Gustafsson Foundation, the Wallenberg Academy Fellows program through the Knut and Alice Wallenberg Foundation, and the European Research Council (ERC) under the European Unions Horizon 2020 research and innovation programme (ERC-2017-StG-757553). 





\begin{thebibliography}{99}
\bibitem{Ali-2018}
Yang, S.-Y., \textit{et.al.}  Symmetry demanded topological nodal-line materials. Adv. Phys.:X \textbf{3,} 1414631 (2018).

\bibitem{Gao-2018}
Gao, H., Venderbos, J. W. F., Kim, Y., and Rappe, A. M. Topological semimetals
from first-principles. 	Annu. Rev. Mater. Res. \textbf{49,} 153-183 (2019).

\bibitem{Chen-2015}
Chen, Y., Lu, Y.-M., and Kee, H.-Y. Topological crystalline metal in orthorhombic perovskite iridates. Nat. Commun. \textbf{6,} 6593
(2015).

\bibitem{Schaffer-2015}
Schaffer, R., Lee, E. K.-H., Lu, Y.-M., and Kim, Y. B. Topological spinon semimetals and gapless boundary states in three dimensions. Phys. Rev. Lett. \textbf{114,} 116803 (2015).

\bibitem{Kim-2015}
Kim, Y., Wieder, B. J., Kane, C. L., and Rappe, A. M. Dirac line nodes in inversion-symmetric crystals. Phys. Rev.
Lett. \textbf{115,} 036806 (2015).

\bibitem{Yu-2015}
Yu, R., \textit{et.al.} Topological node-line semimetal and Dirac semimetal state in antiperovskite Cu$_3$PdN. Phys. Rev. Lett.
\textbf{115,} 036807 (2015).

\bibitem{Bouhon-2017}
Bouhon, A., and Black-Schaffer, A. M. Bulk topology of line-nodal structures protected by space group symmetries in class AI. Preprint at \url{https://arxiv.org/abs/1710.04871} (2017).

\bibitem{PRL.107.186806}
Xu, G., \textit{et.al.} Semimetal and the quantized anomalous Hall effect in HgCr$_2$Se$_4$. Phys. Rev. Lett {\bf 107}, 186806 (2011).
%
\bibitem{PRL.121.106402}
Sun, X.-Q. Zhang, S.-C., and Bzdu\v{s}ek, T. Conversion rules for Weyl points and nodal lines in topological media. Phys. Rev. Lett {\bf 121}, 106402 (2018).
%
\bibitem{Fang-2016}
Fang, C., Weng, H., Dai, X., and Fang, Z. Topological nodal line semimetals. Chin. Phys. B {\bf 25} 117106 (2016).

\bibitem{PRL.121.166802}
Chen, W., Luo, K., Li, L., Zilbergerg, O. Proposal for detecting nodal-line semimetal surface states with resonant spin-flipped reflection. Phys. Rev. Lett. {\bf 121,} 166802 (2018).

\bibitem{Bian-2016-1}
Bian, G., \textit{et. al.} Topological nodal-line Fermions in spin-orbit metal PbTaSe$_2$. Nat. Comm. \textbf{7,} 10556 (2016).

\bibitem{Schoop-2016}
Schoop, L.M., \textit{et. al.} Dirac cone protected by non-symmorphic symmetry and three-dimensional Dirac line node in ZrSiS. Nat. Commun. \textbf{7,} 11696 (2016).


\bibitem{Xie-2015}
Xie, L. S., \textit{et. al.} A new form of Ca$_3$P$_2$ with a ring of Dirac nodes. APL Mater. \textbf{3,} 083602 (2015).

\bibitem{Takane-2018}
Takane, D., \textit{et. al.} Observation of Dirac-like energy band and ring-torus Fermi surface associated with the nodal line in topological insulator CaAgAs. npj Quantum Materials {\bf 3,} 1 (2018).

\bibitem{Bzdusek-2016}
Bzdusek, T., \textit{et.al.} Nodal-chain metals. Nature \textbf{538,} 75-78 (2016).

\bibitem{Weng-2015}
Weng, H., \textit{et. al.} Topological node-line semimetal in three-dimensional graphene networks. Phys. Rev. B \textbf{92,} 045108 (2015).

\bibitem{Pezzini-2017}
Pezzini, S.,\textit{et. al.} Unconventional mass enhancement around the Dirac nodal loop in ZrSiS. Nat. Phys. \textbf{14,} 178-183 (2017).

\bibitem{Wang-2017}
Wang, Y. and Nandkishore, R. M. Topological surface superconductivity in doped Weyl loop materials. Phys. Rev. B {\bf 95,} 060506 (2017).

\bibitem{Nandkishore-2016}
Nandkishore, R. Weyl and Dirac loop superconductors. Phys. Rev. B \textbf{93,} 020506 (2016).

\bibitem{Kopnin-2011}
Kopnin, N. B., Heikkil\"{a}, T. T., and Volovik, G. E. High-temperature surface superconductivity in topological flat-band systems. Phys. Rev. B \textbf{83,} 220503 (2011).

\bibitem{Heikkila-2011}
Heikkil\"{a}, T.T., Kopnin, N.B., and Volovik, G.E. Flat bands in topological media. JETP Lett. \textbf{94,} 233 (2011).

\bibitem{Lothman-2017}
L\"othman, T., and Black-Schaffer, A. M. Universal phase diagrams with superconducting domes for electronic flat bands. Phys. Rev. B {\bf 96,} 064505 (2017).


\bibitem{Sur-2016}
Sur, S. and Nandkishore, R. Instabilities of Weyl loop semimetals. New J. Phys. \textbf{18,} 115006 (2016).


\bibitem{Bergeret-2005}
Bergeret, F. S., Volkov, A. F., and Efetov, K. B. Odd triplet superconductivity and related phenomena in superconductor-ferromagnet structures. Rev. Mod. Phys. {\bf 77,} 1321-1373 (2005).

\bibitem{Berezinski}
Berezinskii, V. L. New model of the anisotropic phase of
superfluid He$^3$. Pisma Zh. Eksp. Teor. Fiz {\bf 20,} 628-631 (1974).

\bibitem{Linder-2017}
Linder, J., Balatsky, A. V. Odd-frequency superconductivity. Rev. Mod. Phys. \textbf{91,} 045005 (2019).

\bibitem{Keizer-2006}
Keizer, R.S., \textit{et.al.} A spin triplet supercurrent through the half-metallic ferromagnet CrO$_2$. Nature {\bf 439,} 825-827 (2006).

\bibitem{Bergeret-2001}
Bergeret, F. S., Volkov, A.F., Efetov, K. B. Long-range proximity effects in superconductor-ferromagnet structures. Phys. Rev. Lett. \textbf{86,} 4096-4099 (2001).

\bibitem{Tanaka-2005}
Tanaka,Y., Asano, Y., Golubov, A. A., and Kashiwaya, S. Anomalous features of the proximity effect in triplet superconductors. Phys. Rev. B \textbf{72,} 140503 (2005).

\bibitem{Black-Schaffer-2013}
Black-Schaffer, A. M., and Balatsky, A. V. Odd-frequency superconducting pairing in multiband superconductors. Phys. Rev. B \textbf{88,} 104514 (2013).

\bibitem{Triola-2017}
Triola, C., and Balatsky, A. V. Pair symmetry conversion in driven multiband superconductors. Phys. Rev. B {\bf 95,} 224518 (2017).

\bibitem{Yokoyama-2011}
Yokoyama, T., Tanaka, Y., and Nagaosa, N. Anomalous Meissner effect in a normal-metal–superconductor junction with a spin-active interface. Phys. Rev. Lett. {\bf 106,} 246601 (2011).

\bibitem{Asano-2011}
Asano, Y., Golubov, A. A., Fominov, Y. V., and Tanaka, Y. Unconventional surface impedance of a normal-metal film covering a spin-triplet superconductor due to odd-frequency Cooper pairs. Phys. Rev. Lett. {\bf 107,} 087001 (2011).

\bibitem{Abrahams-1995}
Abrahams, E., Balatsky, A., Scalapino, D. J., and Schrieffer, J. R. Properties of odd-gap superconductors. Phys. Rev. B \textbf{52,} 1271-1278 (1995).

\bibitem{Asano-2015}
Asano, Y., and Sasaki, A. Odd-frequency Cooper pairs in two-band superconductors and their magnetic response. Phys. Rev. B \textbf{92,} 224508 (2015).


\bibitem{Bernardo-2015b}
Di Bernardo, A., \textit{et.al.} Intrinsic paramagnetic Meissner effect due to $s$-wave odd-frequency superconductivity. Phys. Rev. X {\bf 5,} 041021 (2015).

\bibitem{Alidoust-2014}
Alidoust, M., Halterman, K., and Linder, J. Meissner effect probing of odd-frequency triplet pairing in superconducting spin valves. Phys. Rev. B {\bf 89,} 054508 (2014).

\bibitem{Komendova-2017}
Komendova, L., and Black-Schaffer, A. M. Odd-frequency superconductivity in Sr$_2$RuO$_4$ measured by Kerr rotation. Phys. Rev. Lett. \textbf{119,} 087001 (2017). 

\bibitem{Triola-2018}
Triola C., and Black-Schaffer, A. M. Odd-frequency pairing and Kerr effect in the heavy-fermion superconductor UPt$_3$. Phys. Rev. B {\bf 97,} 064505 (2018).

\bibitem{Eschrig-2003}
Eschrig, M., Kopu, J., Cuevas, J.C., and Schon, G. Theory of half-metal/superconductor heterostructures. Phys. Rev. Lett. {\bf 90,} 137003 (2003).

\bibitem{Asano-2007}
Asano, Y., Tanaka, Y., and Golubov, A. A. Josephson effect due to odd-frequency pairs in diffusive half metals. Phys. Rev. Lett. {\bf 98,} 107002 (2007).

\bibitem{Eschrig-2008}
Eschrig, M., and L\"ofwander, T. Triplet supercurrents in clean and disordered half-metallic ferromagnets. Nat. Phys. {\bf 4,} 138-143 (2008).


\bibitem{Parhizgar-2014}
Parhizgar, F., and Black-Schaffer, A. M. Unconventional proximity-induced superconductivity in bilayer systems. Phys. Rev. B {\bf 90,} 184517 (2014).

\bibitem{Tanaka-1996}
Tanaka, Y., Kashiwaya, S. Theory of the Josephson effect in $d$-wave superconductors. Phys. Rev. B {\bf 53,} R11957-R11960 (1996).

\bibitem{Barash-1996}
Barash, Y.S., Burkhardt, H., Rainer, D. Low-temperature anomaly in the Josephson critical current of junctions in $d$-wave superconductors. Phys. Rev. Lett. {\bf 77,} 4070-4073 (1996).

\bibitem{Tanaka-1997}
Tanaka, Y., Kashiwaya, S. Theory of Josephson effects in anisotropic superconductors. Phys. Rev. B {\bf 56,} 892-912 (1997).

\bibitem{Kwon-2004}
Kwon, H.-J., Sengupta, K., Yakovenko, V.M.  
Fractional ac Josephson effect in $p$- and $d$-wave superconductors
Eur. Phys. J. B {\bf 37,} 349–361 (2004).

\bibitem{Tanaka-2007}
Tanaka, Y., Tanuma, Y., Golubov, A. A.
Odd-frequency pairing in normal-metal/superconductor junctions
Phys. Rev. B {\bf 76,} 054522 (2007).

\bibitem{Tanaka-2007-2}
Tanaka, Y., Golubov, A. A., Kashiwaya, S., Ueda, M. 
Anomalous Josephson effect between even- and odd-Frequency Superconductors.
Phys. Rev. Lett. {\bf 99,} 037005 (2007).

\bibitem{Tanaka-2012}
Tanaka, Y., Sato, M., Nagaosa, N. 
Symmetry and topology in superconductors —odd-frequency pairing and edge states—.
J. Phys. Soc. Jpn. {\bf 81,} 011013 (2012).

\bibitem{RobinsonLinder}
Robinson, J. W. A., Linder, J. 
Strong odd-frequency correlations in fully gapped Zeeman-split superconductors.
Sci. Rep. {\bf 5,} 15483 (2015).

\bibitem{Black-SchafferBalatskyTI}
Black-Schaffer, A. M., Balatsky, A. V.
Odd-frequency superconducting pairing in topological insulators.
Phys. Rev. B {\bf 86,} 144506 (2012).

\bibitem{Black-SchafferBalatskyTI2}
Black-Schaffer, A. M., Balatsky, A. V.
Proximity-induced unconventional superconductivity in topological insulators.
Phys. Rev. B {\bf 87,} 220506(R) (2013).

\bibitem{Black-SchafferBalatskyMultiband}
Black-Schaffer, A. M., Balatsky, A. V.
Odd-frequency superconducting pairing in multiband superconductors.
Phys. Rev. B {\bf 88,} 104514 (2013).

\bibitem{Robinson-2010}
Robinson, J.W.A., Witt, J. D. S. and Blamire, M. G. Controlled injection of spin-triplet supercurrents into a strong ferromagnet. Science {\bf 329,} 59-61 (2010). 

\bibitem{Linder-2015}
Linder J., and Robinson, J. W. A. Superconducting spintronics. Nat. Phys. {\bf 11,} 307-315 (2015).

\bibitem{Bernardo-2015a}
Di Bernardo, A., \textit{et. al.} Signature of magnetic-dependent gapless odd frequency states at superconductor/ferromagnet interfaces. Nat. Commun. {\bf 6,} 8053 (2015).

\bibitem{Cayao-2017}
Cayao, J., and Black-Schaffer, A. M. Odd-frequency superconducting pairing and subgap density of states at the edge of a two-dimensional topological insulator without magnetism. Phys. Rev. B {\bf 96,} 155426 (2017).

\bibitem{Bjornson-2015}
Bj\"ornson, K., and Pershoguba, S. S., and Balatsky, A. V., and Black-Schaffer, A. M. Spin-polarized edge currents and Majorana fermions in one- and two-dimensional topological superconductors. Phys. Rev. B {\bf 92,} 214501 (2015).

\bibitem{Black-Schaffer-2008}
Black-Schaffer, A. M., and Doniach S. Self-consistent solution for proximity effect and Josephson current in ballistic graphene SNS Josephson junctions. Phys. Rev. B {\bf 78,} 024504 (2008).


\end{thebibliography}

\begin{thebibliography}{99}

\bibitem{PRL.107.186806}
Xu, G., \textit{et.al.} Chern semimetal and the quantized anomalous Hall effect in HgCr$_2$Se$_4$. Phys. Rev. Lett {\bf 107}, 186806 (2011).


\bibitem{Tanaka96}
Tanaka, Y., Kashiwaya, S. Theory of the Josephson effect in $d$-wave superconductors. Phys. Rev. B {\bf 53,} R11957-R11960 (1996).

\bibitem{Barash96}
Barash, Y.S., Burkhardt, H., Rainer, D. Low-temperature anomaly in the Josephson critical current of junctions in $d$-wave superconductors. Phys. Rev. Lett. {\bf 77,} 4070-4073 (1996).

\bibitem{Tanaka97}
Tanaka, Y., Kashiwaya, S. Theory of Josephson effects in anisotropic superconductors. Phys. Rev. B {\bf 56,} 892-912 (1997).

\bibitem{Kwon04}
Kwon, H.-J., Sengupta, K., Yakovenko, V.M.  
Fractional ac Josephson effect in $p$- and $d$-wave superconductors
Eur. Phys. J. B {\bf 37,} 349–361 (2004).

\bibitem{Tanaka07}
Tanaka, Y., Tanuma, Y., Golubov, A. A.
Odd-frequency pairing in normal-metal/superconductor junctions
Phys. Rev. B {\bf 76,} 054522 (2007).

\bibitem{Tanaka07-2}
Tanaka, Y., Golubov, A. A., Kashiwaya, S., Ueda, M. 
Anomalous Josephson effect between even- and odd-frequency superconductors.
Phys. Rev. Lett. {\bf 99,} 037005 (2007).

\bibitem{Tanaka12}
Tanaka, Y., Sato, M., Nagaosa, N. 
Symmetry and topology in superconductors —odd-frequency pairing and edge states—.
J. Phys. Soc. Jpn. {\bf 81,} 011013 (2012).

\bibitem{RobinsonLinder}
Robinson, J. W. A., Linder, J. 
Strong odd-frequency correlations in fully gapped Zeeman-split superconductors.
Sci. Rep. {\bf 5,} 15483 (2015).

\bibitem{Black-SchafferBalatskyTI}
Black-Schaffer, A. M., Balatsky, A. V.
Odd-frequency superconducting pairing in topological insulators.
Phys. Rev. B {\bf 86,} 144506 (2012).

\bibitem{Black-SchafferBalatskyTI2}
Black-Schaffer, A. M., Balatsky, A. V.
Proximity-induced unconventional superconductivity in topological insulators.
Phys. Rev. B {\bf 87,} 220506(R) (2013).

\bibitem{Black-SchafferBalatskyMultiband}
Black-Schaffer, A. M., Balatsky, A. V.
Odd-frequency superconducting pairing in multiband superconductors.
Phys. Rev. B {\bf 88,} 104514 (2013).

\end{thebibliography}

\newpage
\section{Supplementary information}

In this supplementary information, we provide additional results to support the findings in the main text. 

\subsection{Effects of anisotropy of the Fermi nodal loop}
As discussed in the main text, a WNL can be modeled with different choices of $\alpha_1$ and $\alpha_2$. For $\alpha_1=\alpha_2$ the Fermi surface at zero doping is almost a circle (a perfect circle if the dispersion is $E \sim k^2$), which we use in the main text. Here we show that our results are not dependent on this particular choice of $\alpha_1$ and $\alpha_2$ and thus the anisotropy of the Fermi nodal loop is not an important factor.
\begin{figure*}[htb]
\includegraphics[width=0.75\linewidth]{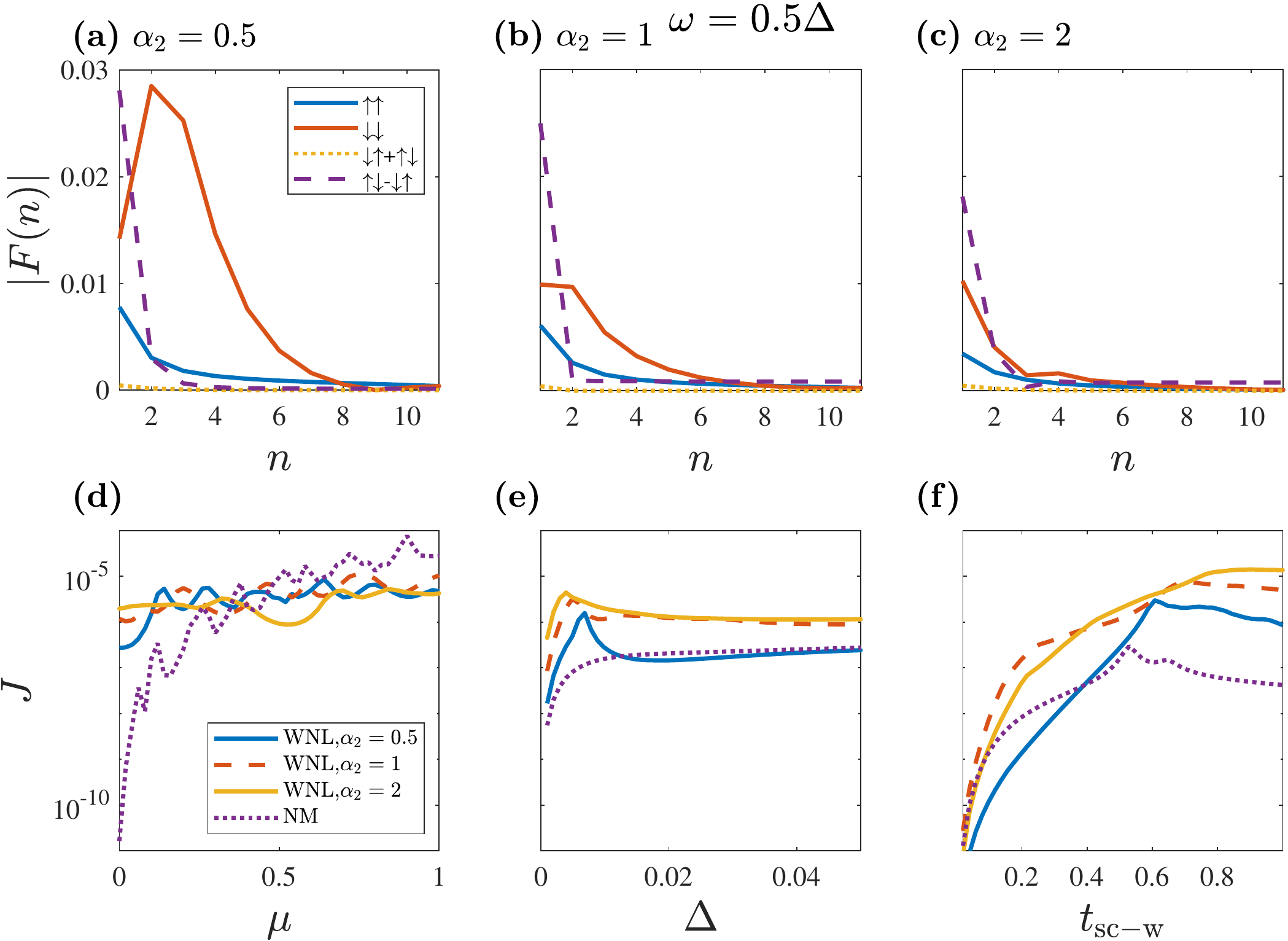}
\caption{Top panels, a-c: Evolution of the absolute value of equal-spin ($\uparrow\uparrow$ and $\downarrow\downarrow$), mixed-
spin ($\uparrow\downarrow+\downarrow\uparrow$) triplet, and spin-singlet ($\uparrow\downarrow-\downarrow\uparrow$) pair amplitudes into the middle of a WNL for $\alpha_2=0.5,1,2$ with $\alpha_1 =1$. All other parameters and model are the same as in Fig.~3(a) in the main text.
Bottom panels, d-f: Josephson current as a function of chemical potential (d), order parameter amplitude $\Delta$ (e), and tunneling between the SC and junction material $t_{\rm sc-w}$ (f). Here we compare the result of the minimal WNL model for $\alpha_2=0.5,1,2$, $\alpha_1=1$ and a NM junction. All other parameters are the same as in Fig.~4 in the main text. }
\label{figS1}
\end{figure*}

In supplementary Fig.~\ref{figS1}(a-c), we set $\alpha_1=1$ and present the propagation of different pair amplitudes through the WNL described in Eq.~(1) in the main text for $\alpha_2=0.5,1,2$, respectively, such that (b) is the same as Fig.~3(a) in the main text. Although the strength of the pair amplitudes changes as we tune $\alpha_2$, we find rapidly decaying spin-singlet and more persistent spin-down triplet pairings throughout the parameter regime. 
In supplementary Fig.~\ref{figS1}(d-f) we then plot the Josephson current as a function of the same parameters as in Fig.~4 in the main text, i.e.~as a function of the chemical potential (d), order parameter $\Delta$ (e), and tunneling between the SC and WNL $t_{\rm sc-w}$ (f), for the same choices of $\alpha_2=0.5,1,2$ and also compare the results with a NM junction. 
As seen in these figures, the large Josephson current in the WNL junction is preserved independently of the anisotropy of the Fermi nodal loop, and the current is always larger or comparable to that of a NM junction, even in the large doping regime. At the lowest doping levels larger $\alpha_2$ creates larger currents, as this term partially governs the hopping in the current direction. However, this is not the only parameter controlling the current and the behavior also changes at higher doping levels.
We also calculate the Josephson current for several other choices of $\alpha_{1,2}$ and find that as long as there exist fully spin-polarized drumhead surface states for the WNL in Eq.~(1) in the main text, the Josephson current is very large in the WNL junction. This further support our claim that the drumhead surface states are crucial for the Josephson effect in WNL junctions.

\subsection{Extended $s$-wave and $p$-wave pair amplitudes}
In the main text we focus on isotropic and $k$-independent pairing, represented by on-site pair amplitudes, but we have also carefully checked for all other commonly present pair amplitudes, up to $p$-wave spatial symmetry. 
First, in supplementary Fig.~\ref{fig:Fwn}, we present the amplitude and frequency dependence of all possible $s$-wave amplitudes residing on nearest neighbor sites in the same layer for the minimal WNL model, Eq.~(1) in the main text. These represent the simplest possible extended-$s$-wave symmetry and can be extracted by taking the summation 
\begin{align}
F(\textbf{r},\omega)=\sum_{k_\parallel}{\rm e}^{{\rm i}k_\parallel\cdot \textbf{r}}F(k_\parallel,\omega),
\end{align}
over all $k_\parallel=(k_x,k_y)$, where $\textbf{r}$ points to the nearest neighbor sites in the square lattice. As the Hamiltonian of the system, Eq.~(1) in the main text, is symmetric and even with respect to $k_x,k_y$, these pairings are the same for all directions of $\textbf{r}$.
supplementary Fig.~\ref{fig:Fwn} is completely analogous to Fig.~2 in the main text, and we see directly that the pair amplitudes are significantly reduced in magnitude; the extended-$s$-wave symmetry amplitudes are only roughly half as large as the isotropic $s$-wave state.
The reduction is actually largest for the spin-singlet pairing, which further emphasizes the importance of the odd-frequency spin-triplet $s$-wave correlations.
\begin{figure*}[htb]
\includegraphics[width=0.9\linewidth]{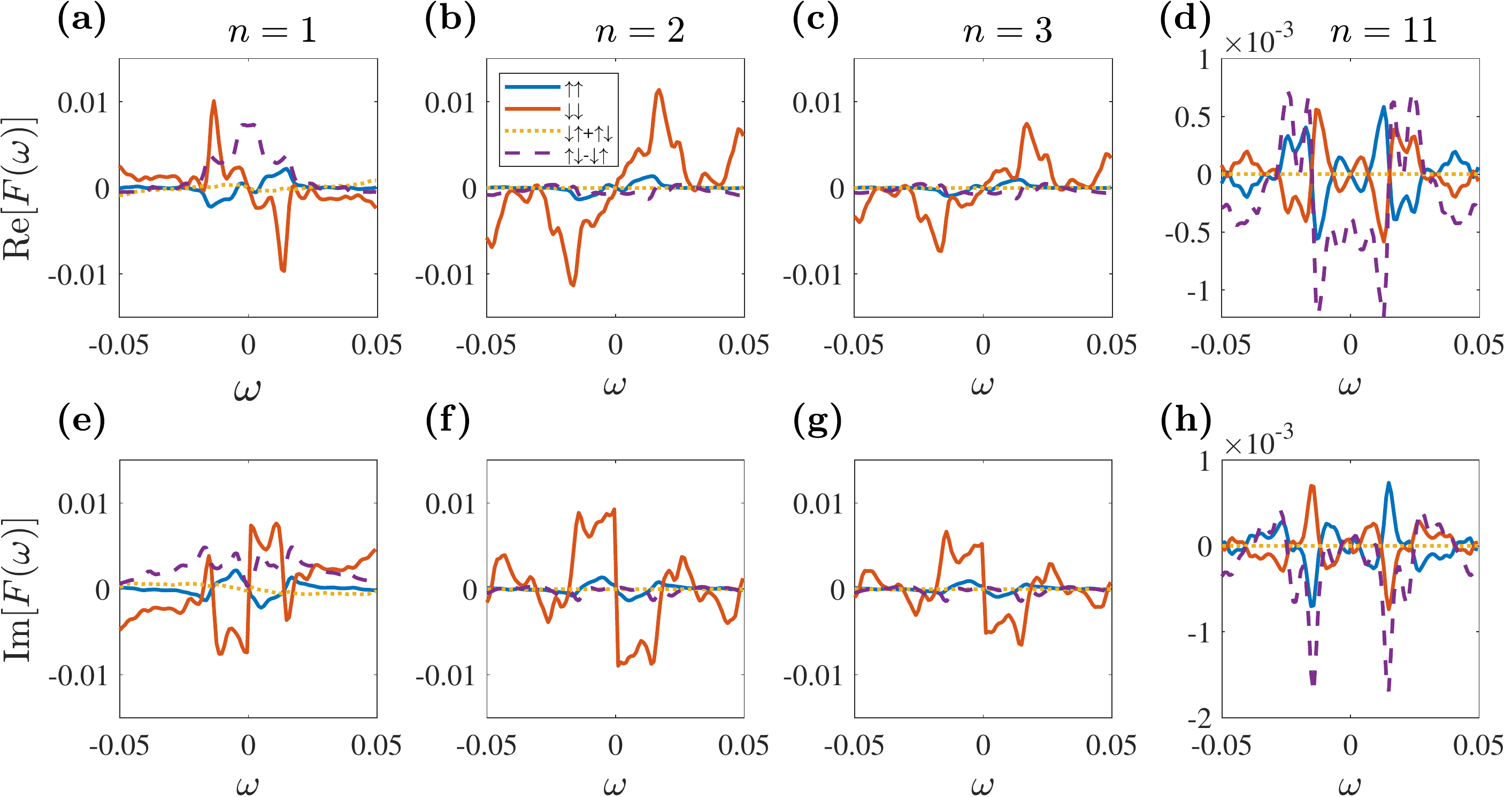}
\caption{Real (top panels, a-d) and imaginary (bottom panels, e-f) parts of the anomalous Green's function $F$ for in-layer extended-$s$-wave symmetry as a function of frequency $\omega$, capturing the pair amplitudes divided into the equal-spin ($\uparrow\uparrow$ and $\downarrow\downarrow$), mixed-spin ($\uparrow\downarrow+\downarrow\uparrow$) triplet, and spin-singlet ($\uparrow\downarrow-\downarrow\uparrow$) components. Left to right figures shows results for the $n=1,2,3$ and middle $n=(n_{\rm w}+1)/2$ layers, respectively. 
Same parameters and model as Fig.~2 in the main text, which present the equivalent plot for the isotropic $s$-wave amplitudes.
}
\label{fig:Fwn}
\end{figure*}

We next consider $p$-wave pairing. This pairing is potentially directly competing with the odd-frequency spin-triplet $s$-wave pairing discussed in the main text (and above), as it has can be of even-frequency and still have spin-triplet symmetry.
Here we first observe that all $p$-wave pairing is completely absent in the $x-y$ plane, in the minimal WNL, as in this plane the Hamiltonian, Eq.~(1) in the main text, possesses even parity and thus proximity effect cannot induce odd-parity pairing. For the generic WNL, Eq.~(2) in the main text, we additionally find that the in-plane $p$-wave pairing is numerically zero.

Left is thus only to consider any possible $p$-wave pairing in-between the layers in the $z$-direction (i.e.~along the direction of propagation of the current) before we can completely exclude that $p$-wave pairing is responsible for the huge Josephson effect in WNLs. 
In supplementary Fig.~\ref{fig4} we show the different $p$-wave pairing amplitudes produced between the first and second layer, i.e.~$\sum_{k_\parallel} F_{12}-F_{21}$, and between the second and third layers, $\sum_{k_\parallel} F_{23}-F_{32}$, for both the minimal and generic WNLs. As expected from a symmetry argument, all spin-triplet $p$-wave amplitudes are even in frequency, while a negligible spin-singlet $p$-wave pairing component is odd in frequency.

For the minimal WNL we find in-between the first and second layers both spin-down triplet and mixed-spin triplet amplitudes, while the rest are all negligible. But already for second-third layer pairing only mixed-spin triplet $p$-wave pairing present. 
Moreover, the magnitude of this mixed-spin triplet state is more than two orders of magnitude smaller than the $s$-wave state in these layers (see Fig.~2 in the main text). This is also expected since the mixed-spin triplet state cannot survive effectively in the spin-polarized normal state.
Thus we can safely conclude that only odd-frequency spin-triplet $s$-wave pairs are of importance inside a minimal WNL Josephson junction.
When instead using the generic WNL, the $p$-wave pairing is slightly altered. Within this model the even-frequency mixed spin-triplet are negligibly small, while the equal spin-triplet pairing components take larger values in comparison with the minimal model. However, these pair amplitudes are still around $2-3$ times smaller than the $s$-wave amplitudes presented in Fig.~2 of the main text. Adding to this the fragility of anisotropic $p$-wave pairs in the presence of disorder and the fact that the minimal and generic WNL junctions carry nearly identical Josephson currents, we can safely conclude that only odd-frequency spin-triplet $s$-wave pairs are of importance inside a WNL Josephson junction.
\begin{figure*}
\includegraphics[width=0.9\linewidth]{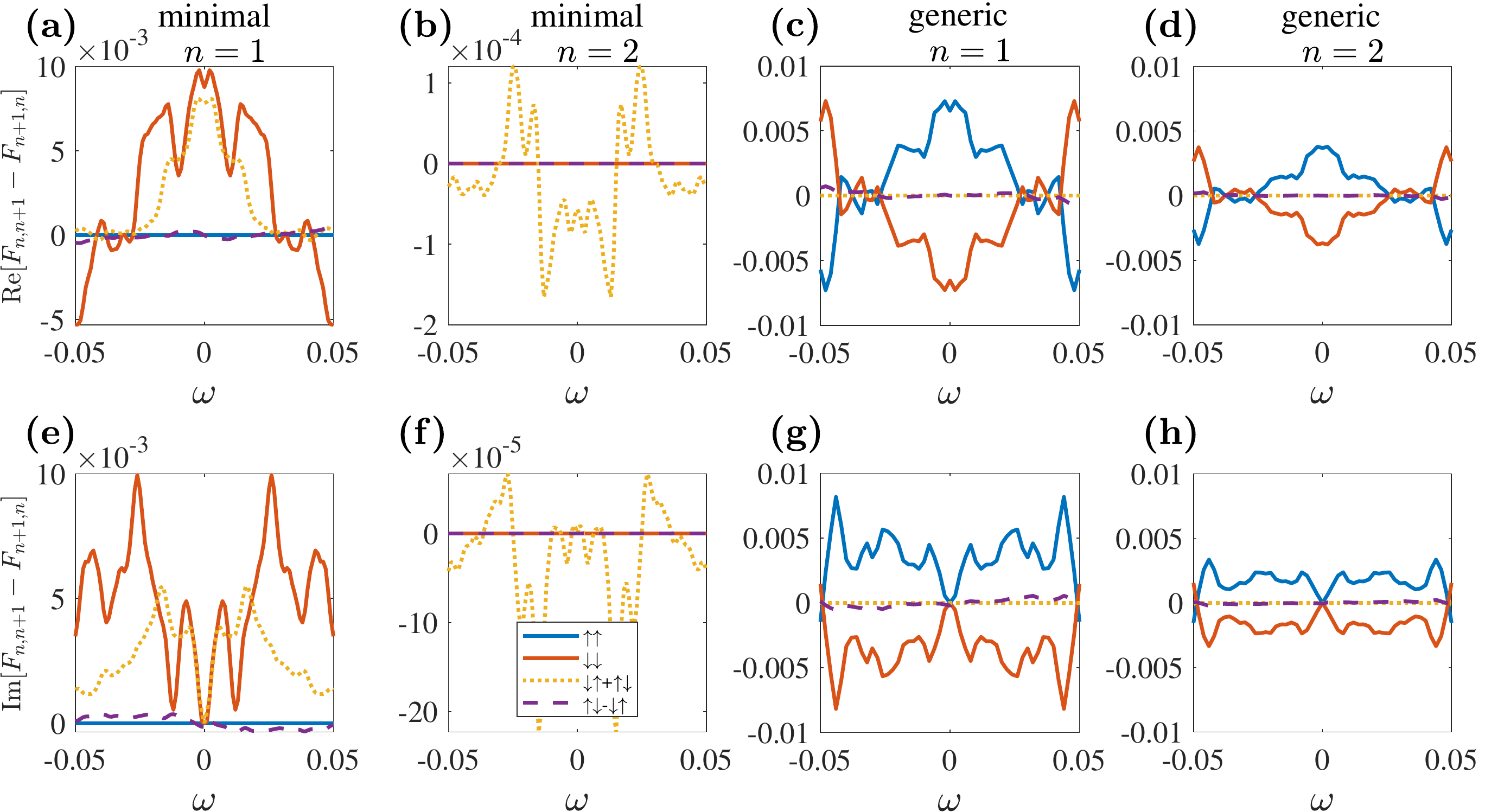}
\caption{Real (top panels, a-d) and imaginary (bottom panels, e-h) parts of anomalous Green's function $F$ for interlayer $p$-wave pairing (the only possible $p$-wave amplitude) as a function of frequency $\omega$, capturing the pair amplitudes divided into the equal-spin ($\uparrow\uparrow$ and $\downarrow\downarrow$), mixed-spin ($\uparrow\downarrow+\downarrow\uparrow$) triplet, and spin-singlet ($\uparrow\downarrow-\downarrow\uparrow$) components. Left to right figures shows results for the first-second layer pairing: $F_{12}-F_{21}$ and second-third layer pairing: $F_{23}-F_{32}$ using minimal and generic models.
Same parameters as in Fig.~2 in the main text, which present the equivalent plot for all on-site $s$-wave amplitudes.}
\label{fig4}
\end{figure*}

\subsection{Frequency dependence of the pair propagation}
In Fig.~3 in the main text we present the absolute value of different pair amplitudes as a function of layer in the WNL sampled at $\omega=0.5\Delta$. Based on this we conclude that the spin-singlet pairing decay extremely rapidly, while the spin-down amplitude survives well in the WNL, effects we attribute to the spin-down spin-polarized drumhead surface states.
In supplementary Fig.~\ref{fig.Fs} we present complementary results for two larger frequencies, $\omega=\Delta$ (a) and $\omega=2\Delta$ (b). The results show that the discussion in the main text focused on $\omega=0.5\Delta$ still holds for larger frequencies, even at energies far above the superconducting gap. In fact, we find that the spin-down triplet amplitudes become even more prominent at larger frequencies, which further underscores the importance of odd-frequency pairing in the WNL.
\begin{figure}[htb]
\includegraphics[width=0.9\linewidth]{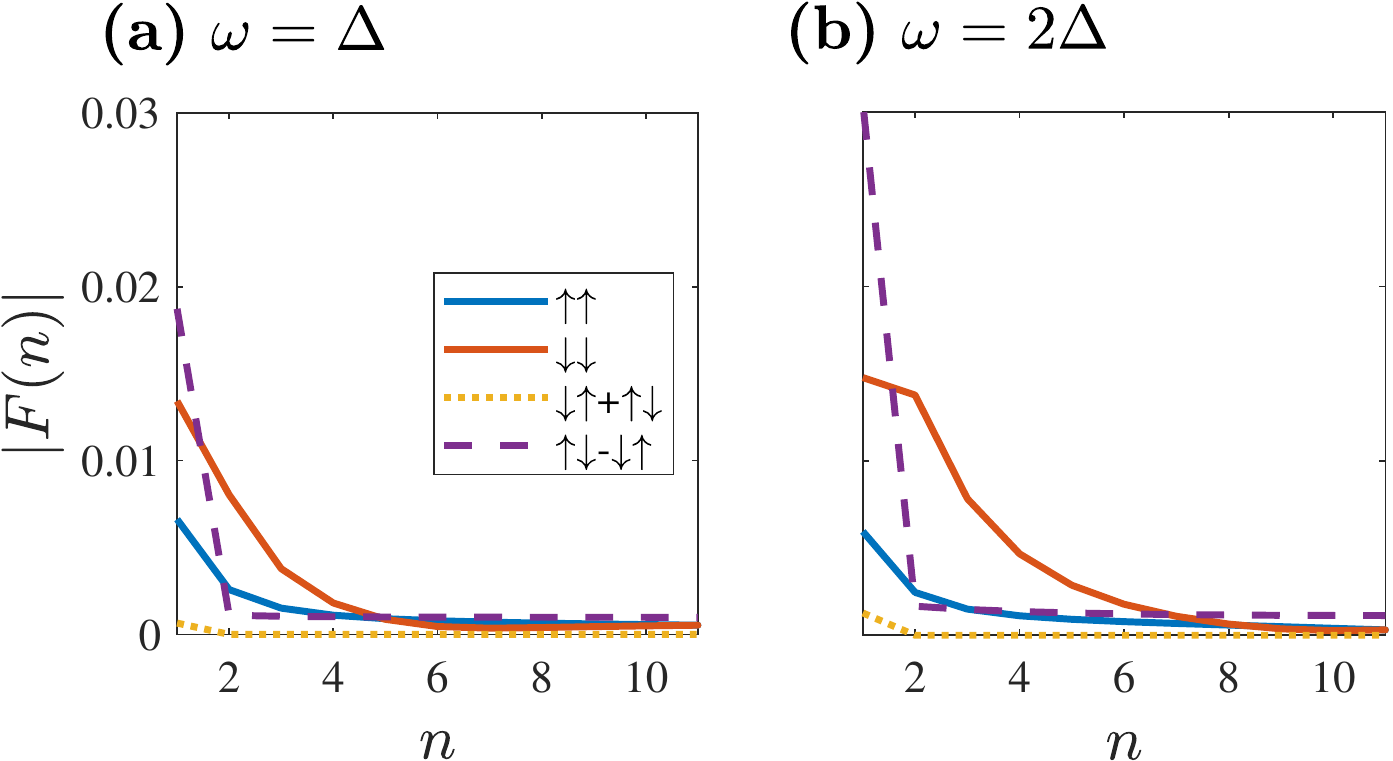}
\caption{Evolution of the absolute value of equal-spin ($\uparrow\uparrow$ and $\downarrow\downarrow$), mixed-spin ($\uparrow\downarrow+\downarrow\uparrow$) triplet and spin-singlet ($\uparrow\downarrow+\downarrow\uparrow$) pair amplitudes into the middle of the WNL extracted at frequencies $\omega=\Delta$ (a) and $\omega=2\Delta$ (b). Same parameters and model as in Fig.~3(a) in the main text, which show the equivalent plot for $\omega = 0.5\Delta$.
}
\label{fig.Fs}
\end{figure}

\subsection{Effect of WNL junction lengths}
In the main text we only give results for a particular thickness of the WNL, $n_{\rm w} =21$. Since the drumhead surface states are very important for the Josephson effect, we might initially think the odd-frequency spin-triplet pairing is also mostly related to a surface effect. After all, it is the spin-polarization of the drumhead state that initially forces the spin-singlet state to diminish in the minimal WNL. 
\begin{figure*}[htb]
\includegraphics[width=0.75\linewidth]{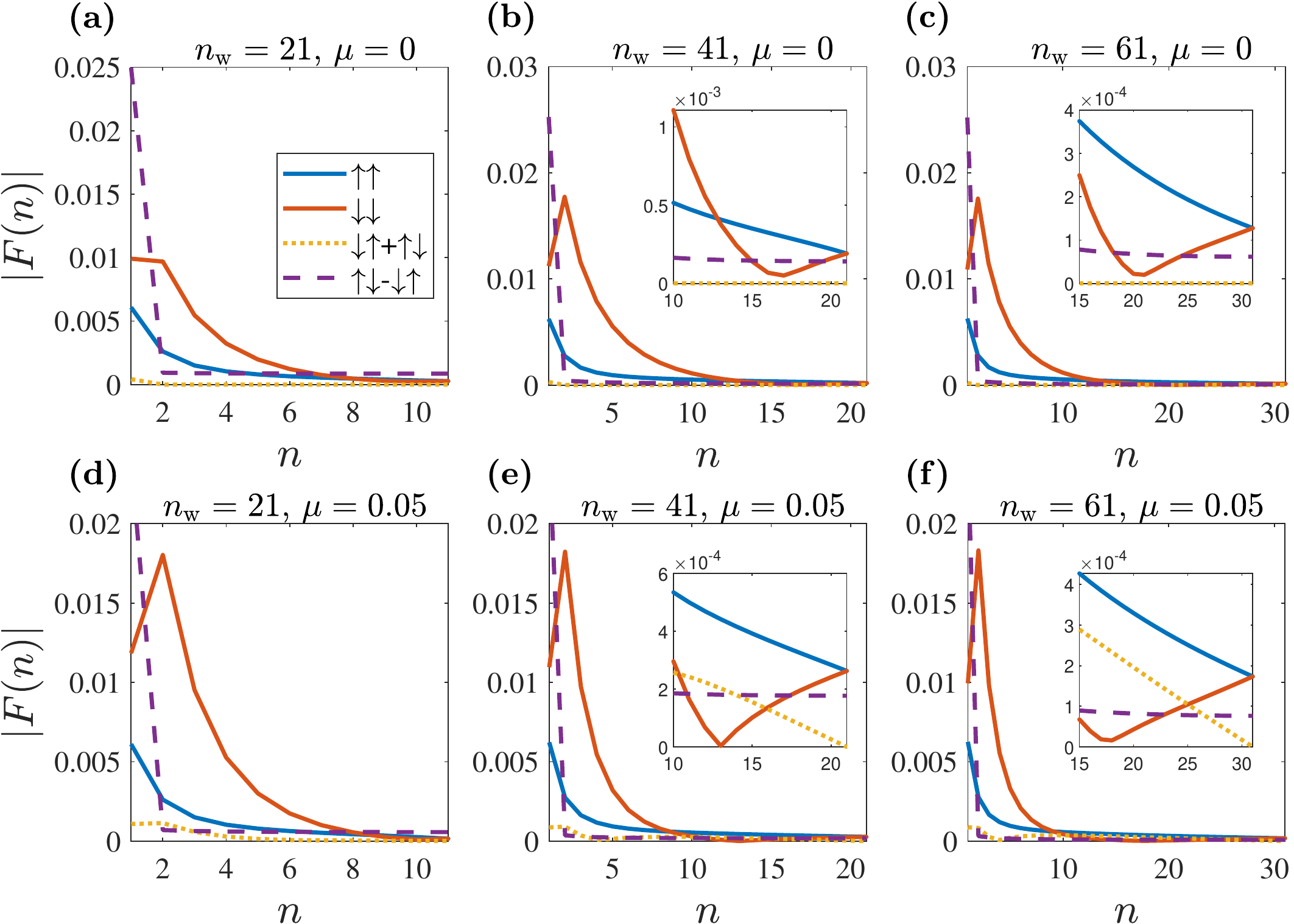}
\caption{Evolution of the absolute value of equal-spin ($\uparrow\uparrow$ and $\downarrow\downarrow$), mixed-spin ($\uparrow\downarrow+\downarrow\uparrow$) triplet and spin-singlet ($\uparrow\downarrow+\downarrow\uparrow$) pair amplitudes into the middle of the WNL extracted at frequency $\omega=0.5\Delta$ for different WNL junction lengths: $n_{\rm w}=21$ (a,d), $n_{\rm w}=41$ (b,e), and $n_{\rm w}=61$ (c,f) and chemical potentials $\mu =0$ (top, a-c) and $\mu = 0.05$ (bottom, d-f). Same parameters and model as in Fig.~3(a) in the main text, which shows the results for for the shortest junction $n_{\rm w} =21$.
}
\label{fig2}
\end{figure*}

In supplementary Fig.~\ref{fig2} we show that this is not correct; odd-frequency superconductivity survives much deeper into the WNL than the conventional spin-singlet pairing. More specifically, we study how the different pairing states propagate into the WNL described by Eq.~(1) in the main text for three different thicknesses: $n_{\rm w}=21,41$ and 61 and also for two different chemical potentials: $\mu=0$ and $0.05$. 
As seen, in the middle of the WNL the spin-triplet pairing is even larger than the spin-singlet pairing for the larger systems $n_{\rm w}=41,61$ than reported in the main text. In fact, for $n_{\rm w} =21$ the spin-singlet state is a bit larger than each individual spin-triplet amplitude in the middle of the WNL, but that is definitely not the case for the longer junctions. We can understand this result by noting that the small spin-singlet component is decaying quickly as the distance from the SC increases while it is the bulk Weyl spin-orbital structure that sustains the odd-frequency spin-triplet pairing. Note that the Weyl dispersion also causes the spin-down pairing to be equal to the spin-up component in the middle of the WNL, and thus both of these contributions should be added to get the full spin-triplet response.

\subsection{Pairing in Dirac nodal loop semimetals}
\begin{figure}[htb]
\includegraphics[width=0.5\linewidth]{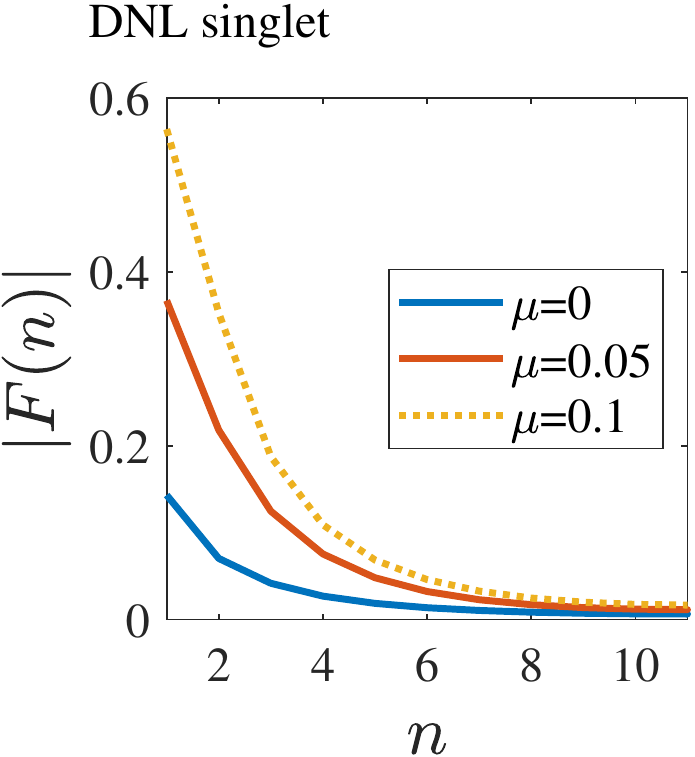}
\caption{Evolution of the absolute value of the spin-singlet ($\uparrow\downarrow-\downarrow\uparrow$) $s$-wave pair amplitude into the middle of the DNL for different chemical potentials. Same parameters and model as in Fig.~3(a) in the main text.
}
\label{fig.Dirac}
\end{figure}

In the main text we state that Dirac Nodal loop (DNL) semimetals, with the same Hamiltonian as the minimal WNL (Eq.~(1) in the main text) but with the Pauli matrices acting in orbital basis instead of spin basis, only hosts spin-singlet pairing due to their spin-degeneracy. Here we provide additional proof and also display how the pair amplitude propagate in the DNL Josephson junction. 
supplementary Fig.~\ref{fig.Dirac} shows the only finite pair amplitude, with even-frequency spin-singlet $s$-wave symmetry, in each layer of the junction and for different chemical potential $\mu=0,0.05,0.1$. All other parameters are set to be the same as Fig.~3 of the main text. 
As the DNL possesses spin-degeneracy, there cannot be any spin-triplet pairing induced by proximity effect from a conventional SC, which we also confirm numerically. Moreover, comparing the spin-singlet $s$-wave pairing of the DNL with that of a NM (see Fig.~3 of main text), it is obvious that the van Hove-liked DOS of the drumhead surface states creates very large pair amplitudes in DNLs. 
For completeness, we have also investigated inter-orbital pairing in the DNL, which, if odd in the orbital index is also odd in frequency. We always find that this pairing is much smaller than the intra-orbital amplitudes reported here and thus there is no effect of odd-frequency pairing in DNL Josephson junctions.

\subsection{Comparison between normal-state DOS in WNL and NM}
The large Josephson current in WNL junctions is particularly remarkable considering the nodal loop or thin torus Fermi surface of the WNL in comparison to a NM. In supplementary Fig.~\ref{fig:Dos} we quantify the statements about the DOS in the WNL in comparison to a NM. Panel (a) shows the DOS in the WNL (minimal model, Eq.~(1) in the main text) as a function of the chemical potential $\mu$. The spin-down polarized surface state is heavily dominating at low $\mu$, while at finite doping, the bulk achieves a comparable DOS. Panel (b) shows the equivalent plot for the the generic WNL (Eq.~(2) in the main text). It shows that both models predict similar surface DOS, although spin-polarization exactly in the surface layer is absent in the generic model. At finite doping $\mu$ the generic WNL also reaches higher DOS in the bulk.
The equivalent plot for the NM is shown in (c). Here the DOS is spin-independent in all layers, has no significant surface contributions, and no singularities at low energies, as expected for the prototypical parabolic dispersion in the NM. The oscillations found in both the WNL and NM densities are due to the finite size of the slabs.
\begin{figure*}[htb]
\includegraphics[width=0.9\linewidth]{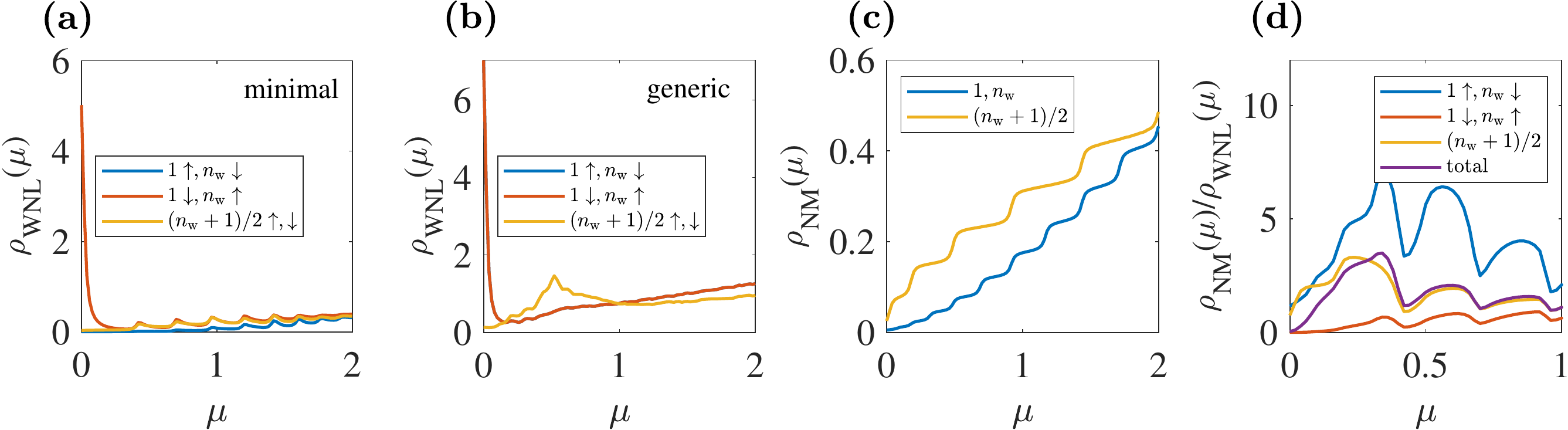}
\caption{DOS of WNL within minimal (a) and generic model (b) together with DOS of NM (c) at zero frequency $\omega=0$ for different layers and spin-polarization, as well as their ratio $\rho_{\rm NM}/\rho_{\rm WNL}$ (d) as a function of chemical potential $\mu$. The total DOS in (d) is the DOS summed over all layers in the slab. Same parameters and models as in Fig.~3(a,d) in the main text, except $\mu$.}
\label{fig:Dos}
\end{figure*}

In supplementary Fig.~\ref{fig:Dos}(d) we directly compare the WNL and NM by plotting the fraction of $\rho_{\rm NM}/\rho_{\rm WNL}$ as a function of $\mu$, divided into spin-polarized surface and bulk contributions.
We see directly that the DOS in the bulk is always higher in the NM compared to the WNL. Thus a NM Josephson junction has more bulk carriers to carry the supercurrent. 
However, in the entire range $\mu\lesssim 0.5$ it is actually the WNL that carries the larger Josephson current, see Fig.~4(a) in the main text. Having chosen the exact same parameters for the SC and the tunneling into the junction for the WNL and NM junctions, it is thus not the bulk DOS that governs the supercurrent. 
The large Josephson effect in lightly doped WNL is instead explained by the drumhead surface states. 
If we also include this DOS in the comparison by summing the DOS in the full slab to arrive at the total DOS, we find that for very small $\mu$ the WNL indeed has the larger total DOS. 
Still, for $0.1 \lesssim \mu \lesssim 0.5$ the total DOS is actually larger in the NM than in the WNL. Thus it is not even the total slab DOS that is important, but also the location of the drumhead states which generates a superior coupling between the external SC contacts and the WNL. This also explains why the two WNL models give nearly identical Josephson currents, since the hosts similar surface DOS.

\subsection{Zero-energy Andreev bound states}
It is rather well-known that the Josephson current in $d$-wave superconductors can be enhanced by surface Andreev bound states \cite{Tanaka96, Barash96, Tanaka97, Kwon04}. It has also been shown that zero-energy Andreev bound states often accompany odd-frequency pairing \cite{Tanaka07, Tanaka07-2, Tanaka12}, although odd-frequency superconductors also exists without zero-energy Andreev bound states \cite{RobinsonLinder, Black-SchafferBalatskyTI, Black-SchafferBalatskyTI2, Black-SchafferBalatskyMultiband}. To investigate the presence of Andreev bound states at the SC-WNL interface we present in supplementary Fig.~\ref{fig3} the local DOS at the four first layers of both a WNL and DNL described by Eq.~(1) in the main text, and compare them with the DOS for the non-superconducting case, i.e.~$\Delta=0$. Here, we set the chemical potential to $\mu=0.1$ to reduce the effect of large density of states of the pristine WNL in the first layers. In both the DNL and WNL the DOS at zero energy for $\Delta=0$ takes an enhanced large value. This large DOS is a surface effect and it decays quickly into the bulk. When turning on superconductivity, a superconducting energy gap is usually induced at the lowest energies. This we also see happening in the DNL, which only hosts conventional proximity effect and where we do not expect any zero-energy Andreev bound states. However, in the WNL we find no suppression at the lowest energies even for a finite superconducting order parameter in the SC leads. This is due to the odd-frequency pairing generating zero-energy Andreev bound states, forcing the DOS to become finite at zero energy. Still, the enhancement in DOS is very small compare to that of the drumhead surface state  (compare with supplementary Fig.~\ref{fig:Dos}(a)), and we therefore conclude that the zero-energy Andreev bound states play a very subdominant role for WNL Josephson junctions.
\begin{figure*}[htb]
\includegraphics[width=0.9\linewidth]{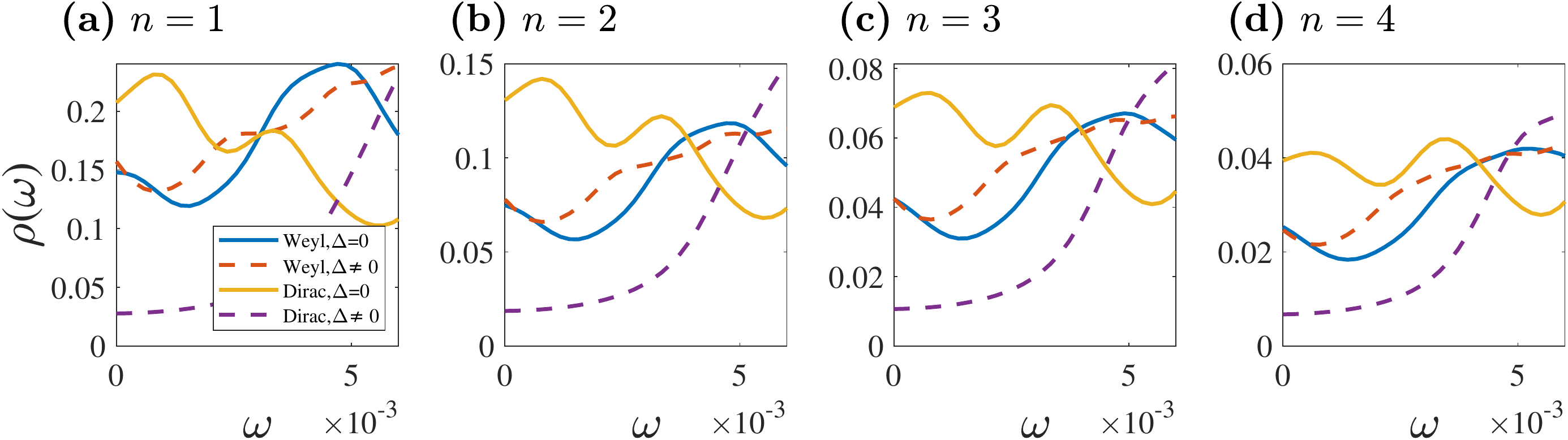}
\caption{DOS of DNL and WNL modeled by Eq.~(1) in the main text and in proximity to a conventional SC with $\Delta=0$ and $\Delta=0.01$ as a function of frequency for layers layers $n=1-4 $(a-d).}
\label{fig3}
\end{figure*}

\subsection{Current-phase relationship}
In an ideal Josephson junction the current-phase relationship is given by $J = J_{\rm max}\sin(\varphi_{\rm L}-\varphi_{\rm R})$. Thus the maximum current is found at $\phi=\varphi_{\rm L}-\varphi_{\rm R} = \pm {\rm \pi}/2$, which is also the value we use in the main text to extract the maximum Josephson current. However, the simple sinusoidal behavior can be modified in real materials and in supplementary Fig.~\ref{fig.Jphi} we explore the full current-phase relationship for WNL Josephson junctions, using Eq.~(1) in the main text. We plot the normalized current $J(\phi)/J_{\rm max}$ for different SC properties, by both varying the order parameter amplitude $\Delta$ and the tunneling between the SCs and WNL $t_{\rm sc-w}$. As seen, $\phi = {\rm \pi}/2$ is an extremely good approximation for generating the largest current for all different parameters, thus supporting this choice in the main text.
\begin{figure}[htb]
\includegraphics[width=0.9\linewidth]{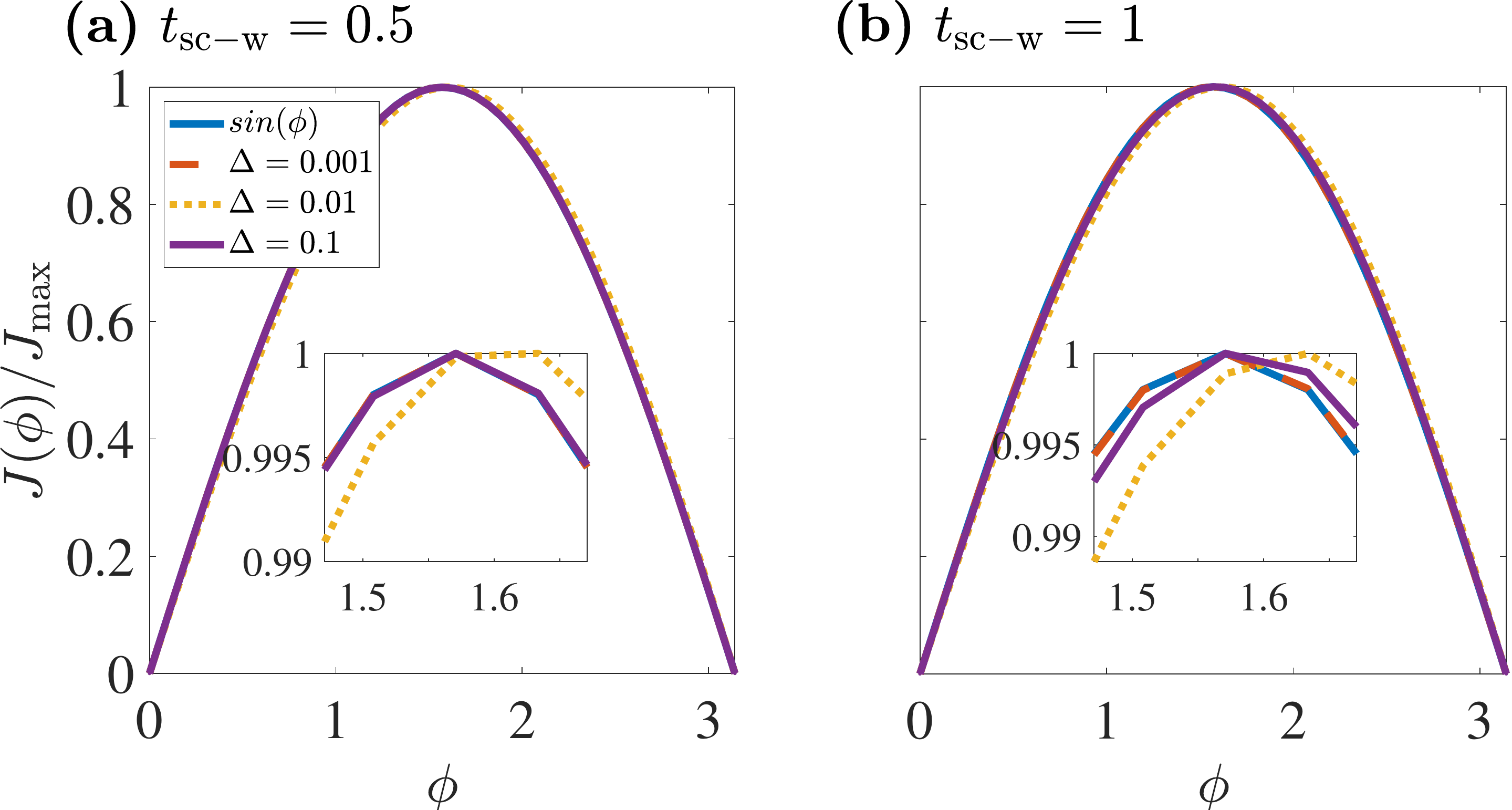}
\caption{Normalized Josephson current through a minimal WNL junction as a function of phase difference between the SCs $\phi=\varphi_{\rm L}-\varphi_{\rm R}$ and compared to the ideal $\sin(\phi)$ relationship. Different curves represent different order parameter amplitudes $\Delta$, with the tunneling between the SCs and WNL set to $t_{\rm sc-w}=0.5$ (a) and $t_{\rm sc-w}=1$ (b). Insets show zoom-in around $\phi = {\rm \pi}/2$. Same parameters and model as in Fig.~2 in the main text, except $\varphi_{\rm L},\varphi_{\rm R}$.}
\label{fig.Jphi}
\end{figure}

\subsection{Currents at other doping levels in NM and HM}
\begin{figure}[htb]
\includegraphics[width=0.9\linewidth]{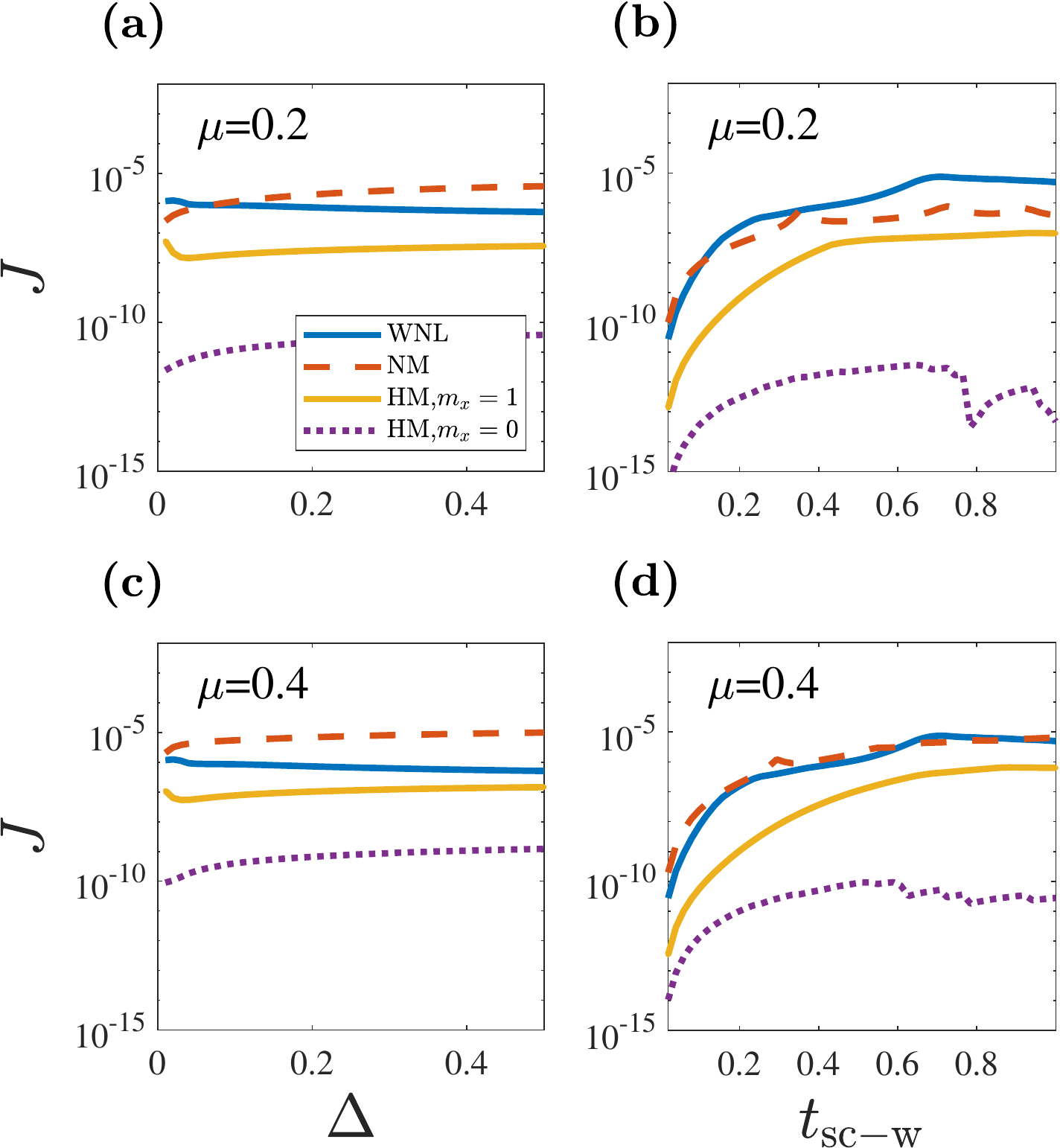}
\caption{Josephson current in a minimal WNL junction at zero doping $\mu_{\rm w} = 0$ in comparison with currents in junctions of NM and HM with and without spin-active region for different doping levels $\mu =0.2, 0.4$ as a function of $\Delta$ (a,c) and $t_{\rm sc-w}$ (a,c). Same parameters and models as in the equivalent Figs.~4(b,c) in the main text, expect the chemical potential $\mu$ in NM and HM. }
\label{fig.Jmun}
\end{figure}
In Fig.~4 in the main text we present the Josephson current for junctions consisting of WNL, NM, and HMs with and without spin-active interface region materials. In Figs.~4(b,c) in the main text we compare the results for different order parameter amplitudes $\Delta$ and tunneling from SCs to the junction material $t_{\rm sc-w}$, respectively, and there have to select a particular doping level. In the main text we use a finite $\mu =0.1$ for the NM and HM junctions in order to get a reasonable but not too large bulk DOS, while for the WNL we use the minimal model in extreme limit of $\mu_{\rm w}=0$. To show that this particular choice of $\mu$ is not misleading, we plot in supplementary Fig.~\ref{fig.Jmun} the equivalent results for $\mu =0.2$ (a,b) and $\mu = 0.4$ (c,d) for the NM and HM, while we keep $\mu_{\rm w} = 0$ in the WNL as a comparison.
For a large range of parameters we find that the WNL Josephson current is larger or of similar magnitude as in the NM junction. It is only for large $\mu$ we find that the NM current surpasses that of the WNL, but not by much and mainly for larger $\Delta$ values. This is driven by an increasing DOS in the NM when $\mu$ is increased. 
For example at $\mu =0.4$ the bulk DOS of the NM is almost five times larger than that of the bulk in the WNL at zero doping. Thus it is still highly remarkable how well the WNL carries a Josephson current compared to the NM.

\subsection{Currents with modified SC properties}
We use prototype conventional spin-singlet $s$-wave SCs as the external contacts in our Josephson junctions. To appropriately capture interface effects we model the SCs with a finite number of layers to make sure bulk conditions are met in the middle of the SCs. 
In supplementary Fig.~\ref{fig.nsc} we show the Josephson current through the minimal WNL material as a function of the thickness of the SCs $n_{\rm sc}$, and also compare it to NM and HM junctions. As seen, there are some oscillatory behavior for small $n_{\rm sc}$ due to finite size effects. However, for $n_{\rm sc} \gtrsim 20$ we approach a nearly constant behavior in all junctions. Thus the choice of $n_{\rm sc} = 20$ in the main text is a very good compromise between studying a small system for computational purposes and reaching good bulk conditions in the SC contacts.
Moreover, we also test our main results for different values of the chemical potential $\mu_{\rm SC}$ in the SC  and the interface tunneling $t_{\rm sc}$. We find that for all larger values of the chemical potential, i.e.~modeling a good metallic normal state as expected for a conventional SC, the results do not vary significantly and our choice of $\mu_{\rm sc}=2$ in the main text is very representative. Choosing different tunneling $t_{\rm sc}$ also do not qualitatively change the result and our main conclusions holds for different models of the SC.
\begin{figure}[htb]
\includegraphics[width=0.5\linewidth]{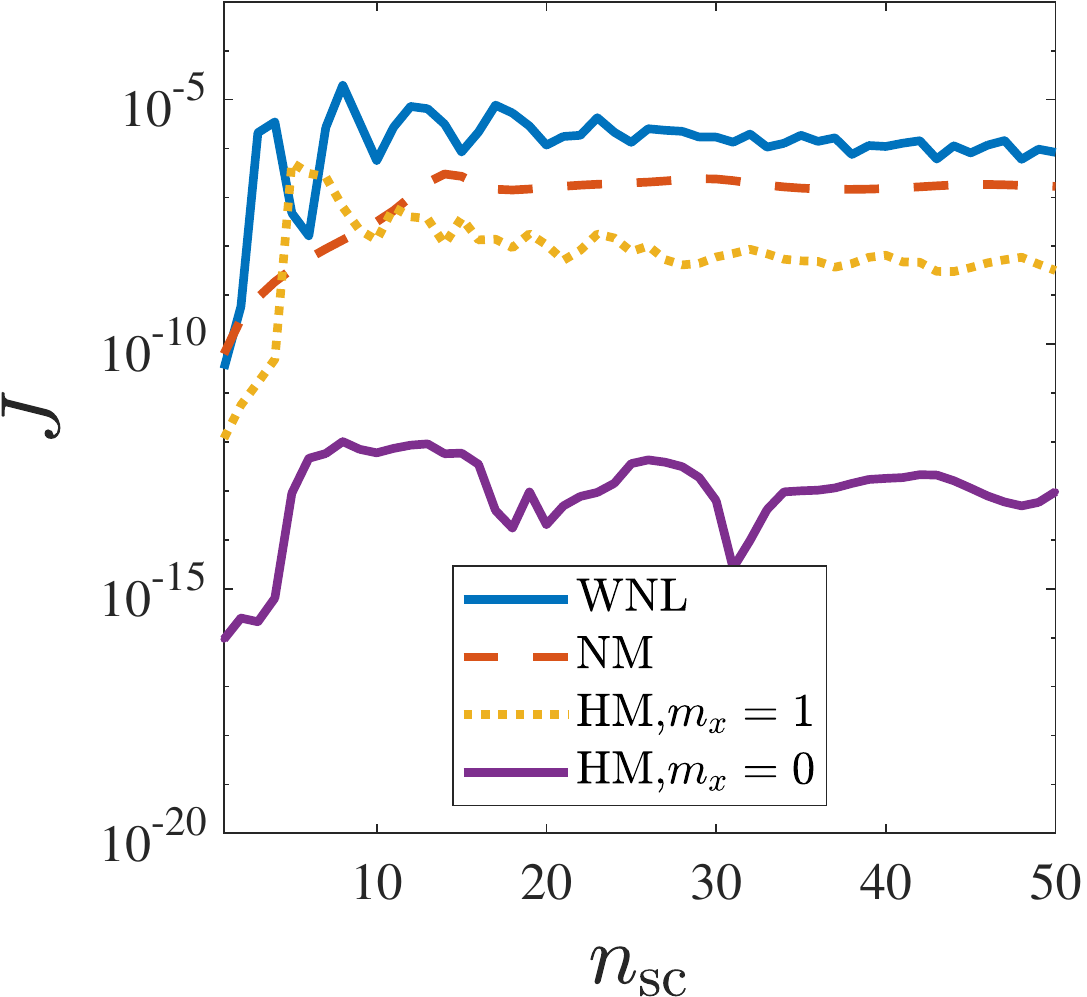}
\caption{Josephson current in WNL, NM, and HM junctions as a function of the thickness of the SC contacts, $n_{\rm sc}$. Same parameters and models as in Figs.~3(a,d,e,f) in the main text.}
\label{fig.nsc}
\end{figure}

\end{document}